\documentclass[pdflatex,sn-mathphys-ay]{sn-jnl}

\usepackage{graphicx}%
\usepackage{multirow}%
\usepackage{amsmath,amssymb,amsfonts}%
\usepackage{amsthm}%
\usepackage{mathrsfs}%
\usepackage[title]{appendix}%
\usepackage{xcolor}%
\usepackage{textcomp}%
\usepackage{manyfoot}%
\usepackage{booktabs}%
\usepackage{algorithm}%
\usepackage{algorithmicx}%
\usepackage{algpseudocode}%
\usepackage{listings}%
\usepackage{caption} 

\theoremstyle{thmstyleone}%
\theoremstyle{thmstyletwo}%

\theoremstyle{thmstylethree}%

\begin{document}

\title[Article Title]{A Comprehensive Approach to Carbon Dioxide Emission Analysis in High Human Development Index Countries using Statistical and Machine Learning Techniques}


\author{\fnm{Hamed} \sur{Khosravi}}\email{hamed.khosravi@mail.wvu.edu}

\author{\fnm{Ahmed Shoyeb} \sur{Raihan}}\email{ar00065@mix.wvu.edu}

\author{\fnm{Farzana} \sur{Islam}}\email{fi00003@mix.wvu.edu}

\author{\fnm{Ashish} \sur{Nimbarte}}\email{Ashish.Nimbarte@mail.wvu.edu}

\author*{\fnm{Imtiaz} \sur{Ahmed}}\email{imtiaz.ahmed@mail.wvu.edu}

\affil{\orgdiv{Department of Industrial and Management Systems Engineering}, \orgname{West Virginia University}, \orgaddress{\city{Morgantown}, \postcode{26505}, \state{West Virginia}, \country{USA}}}


\abstract{Reducing Carbon dioxide (CO2) emission is vital at both global and national levels, given their significant role in exacerbating climate change. CO2 emission, stemming from a variety of industrial and economic activities, are major contributors to the greenhouse effect and global warming, posing substantial obstacles in addressing climate issues. It's imperative to forecast CO2 emission trends and classify countries based on their emission patterns to effectively mitigate worldwide carbon emission. This paper presents an in-depth comparative study on the determinants of CO2 emission in twenty countries with high Human Development Index (HDI), exploring factors related to economy, environment, energy use, and renewable resources over a span of 25 years. The study unfolds in two distinct phases: initially, statistical techniques such as Ordinary Least Squares (OLS), fixed effects, and random effects models are applied to pinpoint significant determinants of CO2 emission. Following this, the study leverages supervised and unsupervised machine learning (ML) methods to further scrutinize and understand the factors influencing CO2 emission. Seasonal AutoRegressive Integrated Moving Average with eXogenous variables (SARIMAX), a supervised ML model, is first used to predict emission trends from historical data, offering practical insights for policy formulation. Subsequently, Dynamic Time Warping (DTW), an unsupervised learning approach, is used to group countries by similar emission patterns. The dual-phase approach utilized in this study significantly improves the accuracy of CO2 emission predictions while also providing a deeper insight into global emission trends. By adopting this thorough analytical framework, nations can develop more focused and effective carbon reduction policies, playing a vital role in the global initiative to combat climate change.}




\maketitle

\section{Introduction}\label{sec1}

Climate change stands as one of the most formidable challenges of our time, widely recognized as an unequivocal and potentially irreversible threat to the sustainability of our global environment \citep{hao2016influence}. The consensus among scientists and researchers is overwhelming, with the Intergovernmental Panel on Climate Change (IPCC) explicitly stating the critical impacts and long-term implications of climate change on ecosystems, biodiversity, and human societies \citep{pettorelli2021time}. At the heart of this environmental conundrum lies the pivotal role of CO2  emission, identified as a primary contributor to the greenhouse effect and a key driver of global warming \citep{liu2019causes, zheng2019review}. CO2 emission result predominantly from the burning of fossil fuels, such as coal, oil, and natural gas, a practice intricately linked to industrial activities, energy production, and transportation \citep{gur2022carbon, yoro2020co2}. The accumulation of CO2 in the Earth’s atmosphere traps heat, leading to a rise in global temperatures, a phenomenon commonly referred to as global warming. The effects of this warming are far-reaching, ranging from more frequent and severe weather events, such as hurricanes and droughts, to long-term shifts in climate patterns, impacting agriculture, ecosystems, and sea levels worldwide \citep{loucks2021impacts}. Recent research has further solidified the understanding of CO2’s impact on climate change. 

Globally, countries are implementing diverse strategies and policies to combat the seriousness of climate change, such as transitioning to renewable energy sources, enhancing energy efficiency, and engaging in international agreements like the Paris Agreement \citep{fekete2021review, seo2017beyond}. Efforts like the REDD+ (Reducing emission from deforestation and forest degradation in developing countries) programs focus on reducing emission from deforestation and forest degradation, recognized for their cost-effectiveness in mitigating greenhouse gases \citep{rakatama2017costs}. These approaches, ranging from carbon pricing mechanisms to investments in green technologies, underscore the importance of understanding the various factors influencing CO2 emission. This understanding is crucial for tailoring specific, effective strategies, anticipating, and mitigating unintended policy consequences, facilitating informed international cooperation, and ensuring the adaptability and longevity of climate change mitigation efforts. In essence, the success of global initiatives to reduce carbon emission hinges on a nuanced comprehension of the determinants of CO2 emission, a cornerstone in designing and implementing impactful environmental policies \citep{cai2018nexus, li2022impacts, puertas2021eco}. In the design and implementation of such programs, a critical element is the identification of factors affecting CO2 emission. This requires a multifaceted approach, considering not only direct factors like deforestation and fossil fuel use but also indirect drivers such as economic policies, energy consumption patterns, technological advancement, and societal behaviors \citep{jiang2018investigating}. A detailed understanding of these factors is essential for developing effective strategies that can significantly reduce emission \citep{wang2023novel}.

This research paper seeks to unravel the complexities surrounding CO2 emission, with a specific focus on countries with high Human Development Index (HDI) scores \citep{adekoya2021renewable, hossain2021nexus}. These countries, often seen as the vanguards of technological advancement and economic prosperity, find themselves at a critical juncture, where their developmental strides are paradoxically intertwined with substantial environmental footprints \citep{mohmmed2019driving}. The significance of understanding CO2 emission in high-HDI countries cannot be overstated. These nations, which have historically powered their growth through extensive use of fossil fuels, are now at the forefront of a global crisis they significantly contributed to \citep{ahmad2020critical}. The environmental policies and technological innovations that these countries adopt or ignore will have far-reaching consequences, not only within their borders but across the globe. Thus, dissecting the fabric of CO2 emission in these contexts becomes not only an environmental imperative but also a moral one. The rationale for focusing on high-HDI countries in this study is three-fold. Firstly, these countries have been the primary architects of the current state of global CO2 emission \citep{mardani2020multi}. Their industrial revolutions, fueled by carbon-intensive resources, have left an indelible mark on the planet's atmosphere. Understanding their emission patterns is crucial to untangling the historical context of the current climate crisis and in charting a course for global emission reduction strategies. Secondly, high-HDI countries are often at the cutting edge of technological innovation and economic transformation \citep{le2019drivers}. This positions them uniquely to lead the way in developing and implementing solutions to reduce CO2 emission. From renewable energy technologies to policy innovations for carbon reduction, these countries have the potential to set precedents and create models that can be emulated worldwide. Thirdly, the economic structures and lifestyles in high-HDI countries significantly influence global consumption patterns and, by extension, CO2 emission \citep{hao2022effect}. These nations, therefore, have a disproportionate impact on setting global trends. Their commitment to sustainable practices, or lack thereof, can either exacerbate the climate crisis or steer the world towards a more sustainable trajectory. This research aims to contribute to this vital global endeavor by providing a nuanced understanding of the determinants of CO2 emission in high-HDI countries, thereby laying the groundwork for informed, effective, and sustainable policy interventions.

The intricate relationship between CO2 emission and various socio-economic, environmental, and energy-related factors has been a focal point of numerous studies over the past decades \citep{fan2020empirical, rahman2020disaggregated, xie2020does, zhang2020examining}. These research endeavors have collectively contributed to a growing understanding of the multifaceted nature of CO2 emission. From an economic perspective, a substantial body of research has explored the relationship between emission and variables like GDP growth, levels of industrialization, and urbanization trends. \citep{ding2017examining, li2016ghg, sikder2022integrated}. Environmental studies have explored the impact of land use changes, deforestation, and climate policies \citep{arneth2017historical, popp2012additional, raihan2022dynamic}. In terms of energy factors, the role of energy consumption, the mix of energy sources, and the efficiency of energy use have been thoroughly scrutinized \citep{bilgili2023research, nejat2015global, shah2023role}. Despite the breadth of the existing research, there still remain notable gaps and limitations. One main limitation is the predominant focus on cross-sectional data analysis, which often fails to capture the dynamic nature of the relationship between CO2 emission and their determinants over time \citep{hashmi2019dynamic}. Additionally, many studies have been constrained by the scope of their geographical focus, often limiting their analysis to a single country or a small group of countries, thereby restricting the generalizability of their findings \citep{ahmadi2023carbon, fang2018novel, li2023prediction, wang2020modeling}. Another notable gap is the insufficient exploration of the interaction effects between different determinants of CO2 emission, a complexity that demands a more nuanced analytical approach \citep{adebayo2021determinants, dong2019determinants, zhou2018examining}.

By employing a combination of advanced econometric models and innovative analysis techniques, this study aims to bridge the existing gaps and provides the following contributions.

\begin{figure}[!htb]
\centering
\includegraphics[width=0.95\linewidth]{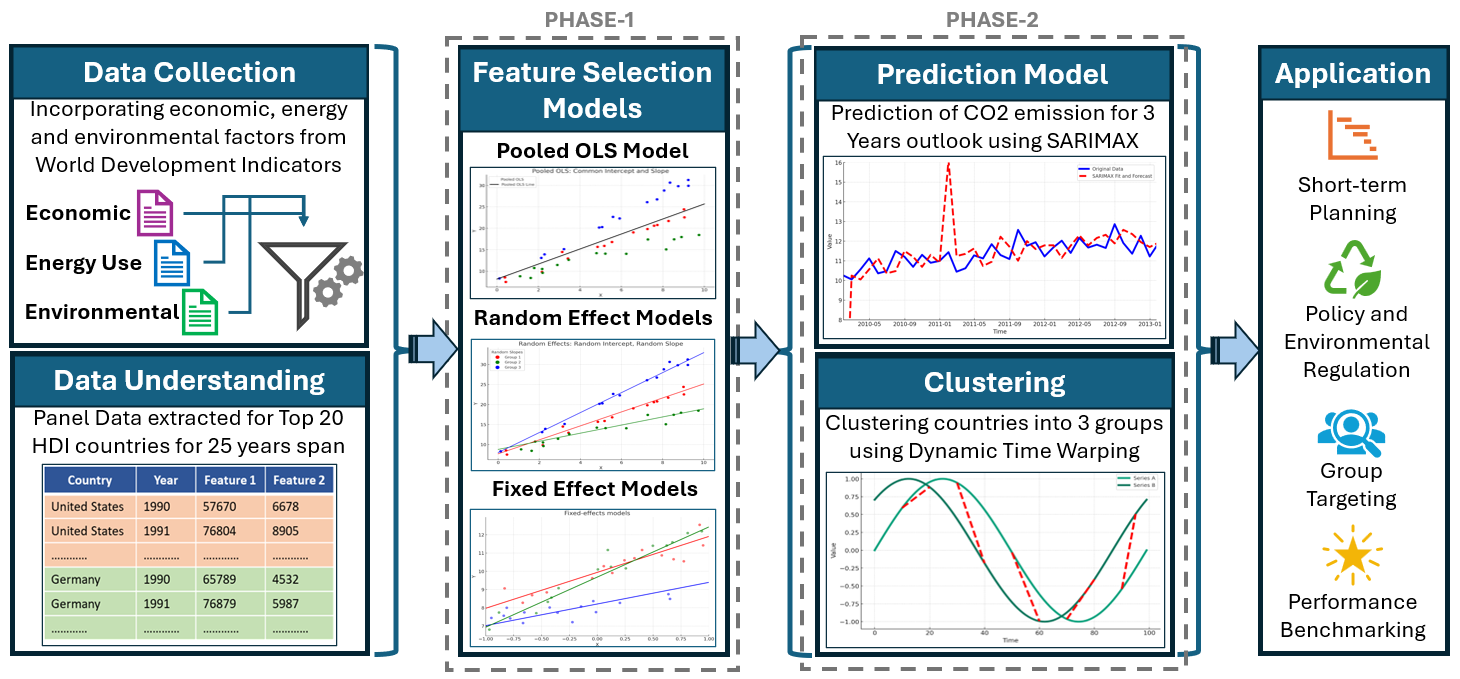}
\caption{The overview of the utilized framework}
\label{fig:myimage1}
\end{figure}

\begin{itemize}
    \item We consider a wide variety of features spanning economic, environmental, energy use, electricity consumption, and renewable energy dimensions, ensuring a comprehensive representation of potential determinants of CO2 emission.
    \item We analyze data for 20 high HDI countries, covering a span of 25 years, providing a significant temporal and spatial scope for the study. The data is exported from a reputable online source, ensuring the reliability and quality of the dataset used for analysis. The study of CO2 emission within high-HDI countries offers not just an insight into the emission themselves but also into the broader narrative of environmental responsibility and sustainability. It is within these nations that the battle against climate change could be most effectively strategized and executed.
    \item We employ five different statistical models utilizing Ordinary Least Squares (OLS), fixed effects, and random effects based models on the panel dataset. Panel data combines elements of both cross-sectional data and time series data which tracks multiple subjects such as countries across different points in time. This varied approach in modeling has allowed us to conduct a comprehensive comparative analysis, crucial for pinpointing the key predictive features. Additionally, the diversity of these models enhanced our ability to effectively analyze cross-sectional time series data, providing a deeper understanding of temporal trends and variations across different countries.
    \item We adopt a structured, two-phase approach. In the initial phase, we focus on identifying the significance of different variables that influence CO2 emission. The subsequent phase aims at forecasting CO2 emission for the forthcoming three years and groups countries into clusters according to their emission trajectories. This clustering plays a pivotal role in deciphering and classifying the diverse emission trends across nations, offering critical insights that can inform the creation of precise environmental policies and foster international collaboration. 

\end{itemize}

An illustration of our comprehensive two-stage framework is presented in Figure \ref{fig:myimage1}. The numerical study has successfully validated our forecasting models, achieving a higher accuracy in predicting CO2 emission and closely mirroring actual trends. This highlights the proposed framework's effectiveness in providing dependable emission projections for the near future.

The rest of the paper is organized as follows. Section 2 provides a comprehensive review of the literature. Section 3 presents the panel dataset used in this study. Section 4 describes the preliminaries, definitions, and steps of our proposed method. Section 5 highlights simulation, computational results, and the effectiveness of our method. Finally, Section 6 provides concluding remarks and suggestions for future research.

\section{Literature Review}\label{sec2}

\begin{figure}[!htb]
\centering
\includegraphics[width=0.8\linewidth]{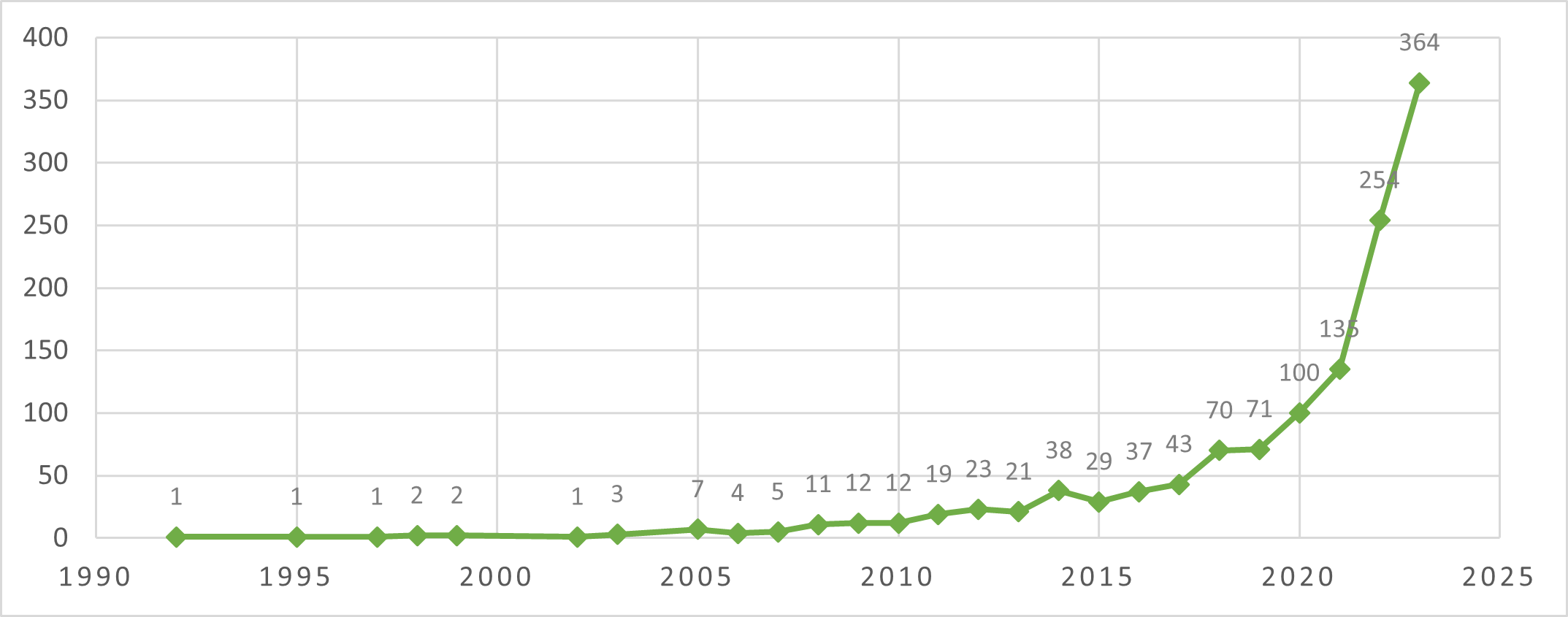}
\caption{Trend of research in carbon emission prediction from an analysis of 1266 documents}
\label{fig:myimage2}
\end{figure}

The increasing focus on carbon emission prediction is a response to the urgent need to address climate change and its impacts \citep{jiang2019research}. Figure~\ref{fig:myimage2} illustrates a marked rise in carbon emission prediction research, showing an upswing in publications from 2010 onwards. This trend underscores an increasing academic and scientific interest in comprehending and forecasting carbon emission, driven by escalating concerns over global climate change. As the primary driver of global warming, CO2 emission are a key focus in efforts to mitigate climate change \citep{zheng2019review}. Accurate predictions are essential for informing and guiding policy decisions, developing effective strategies for emission reduction, and meeting international climate targets, such as those outlined in the Paris Agreement \citep{van2021net}. Additionally, these studies play a vital role in advancing our understanding of the complex interplay between economic development, energy use, technological advancement, and environmental sustainability. By improving the accuracy and reliability of emission forecasts, this growing body of research contributes significantly to global efforts to protect the environment and ensure a sustainable future \citep{olabi2022assessment}. The extant literature on carbon emission prediction, comprising investigations that pinpoint critical determinants and guide policy formulation, can be broadly categorized into two distinct streams: one consists of research dedicated to a single country or various regions within a single nation, while the other encompasses studies that concurrently address carbon emission across multiple countries. 

The United States is a leader in climate research, supported by top-tier universities and institutions specializing in carbon emission prediction, propelling it to a prominent role in efforts to understand and mitigate climate change \citep{bennedsen2021modeling, wang2020decomposition}. China's increasing focus on carbon management research is driven by its significant economic growth and large population, reflecting its crucial role in global environmental sustainability \citep{tian2022research, wang2022forecasting}. Germany is recognized for its dedication to renewable energy and climate technology, contributing significantly to climate change and emission prediction research \citep{huber2021carbon}. The United Kingdom, with its extensive research centers, plays a vital role in advancing climate science and policy \citep{adedoyin2020energy}. Australia's research efforts, motivated by unique environmental challenges, contribute to both national and international carbon management strategies \citep{sarkodie2018assessment}. Canada's emphasis on sustainable development and the integration of environmental, economic, and social factors distinguishes its emission prediction research \citep{javanmard2023forecast}. Japan leverages its advanced technology sector to address its vulnerability to climate change, focusing on the intersection of technology, policy, and environmental science \citep{jiang2020carbon}. Post-Paris Agreement, France leads in developing innovative carbon emission reduction strategies \citep{dong2018probability}. India's research is increasingly important, addressing the challenges and opportunities presented by its rapid development and large population \citep{bakir2022forecasting}. The Scandinavian countries, known for their commitment to sustainability, provide insightful research on emission, serving as models of sustainability \citep{ghalandari2021energy}. While there are several country-specific studies, the literature also is enriched with studies focusing on multiple countries. For instance, several studies have forecasted CO2 emission and analyzed the most influential factors behind the emission by conducting studies on the G20 countries \citep{alotaibi2021using, mardani2018energy, zhang2023novel}. The OECD, or Organization for Economic Co-operation and Development, is a larger group of countries which is another area of carbon emission research \citep{calbick2014differences, pan2019influential, wang2021impacts}. Besides, multiple studies have been conducted based on European countries, Middle-eastern countries, Asian countries and so on \citep{ahmadi2023carbon, marotta2023greenhouse, sasaki2021predicting}. While many researchers have also focused on developed countries, countries with emerging economies, others have grouped multiple countries together based on different criteria such as geographical locations, political relations, and so on. 

Clustering countries based on their CO2 emission profiles emerges as a pivotal strategy, offering a tailored approach to environmental policy and action. The spatial clustering of industrial CO2 emission efficiency across China's provinces underscores a critical need for targeted regional policy support \citep{zhang2016co2}. Such targeted approaches necessitate the adoption of innovative clustering methodologies to dissect the intricate relationship between CO2 emission and economic growth globally. This understanding paves the way for the formulation of tailored policy interventions, uniquely suited to the diverse developmental trajectories encountered across countries, thereby highlighting the importance of nuanced environmental strategies \citep{li2019cluster}. Further emphasizing the role of informed policymaking, the Kyoto Protocol has not considered its effects on agriculture and forestry, which further confirms that the clustering of countries can aid in more tailored policies \citep{gallo2018clustering}. Similarly, the European Union's emission Trading System exemplifies the efficacy of market-based mechanisms. By setting a precedent for global climate action, it showcases how strategic market interventions can lead to significant reductions in greenhouse gas emission, underscoring the need for strategic and tailored policies \citep{lamb2022countries}. The detailed analysis of CO2 emission determinants within Far East countries highlights the urgent need for policies that promote green urbanization, strategic industrial regulation, and the increased adoption of renewable energy. It also shows that the clustering of countries can assist in creating tailored policies for urgent action in specific regions \citep{anwar2020impact}. Additionally, insights gained from structural decomposition \citep{liu2019causes} analysis to  examine the varying impacts observed across different country groups calls for a sophisticated and multifaceted approach to global warming. This approach must acknowledge and address the diverse factors influencing emission. The necessity of clustering countries based on their CO2 emission profiles and reduction capacities becomes evident when considering the collective insights from these analyses. By categorizing countries into clusters that reflect their unique emission characteristics and developmental challenges, policymakers and international bodies can devise more effective, context-specific strategies for CO2 emission control.

Building upon the existing literature on carbon emission prediction, an extensive array of research has been conducted to identify and evaluate the effectiveness of various indicators that influence carbon emission. These indicators, ranging from economic factors such as GDP and industrial activity to energy consumption metrics and technological advancements, are paramount in constructing accurate predictive models \citep{lin2023greenhouse}. Economic indicators have been extensively studied to understand their relationship with carbon emission. Gross Domestic Product (GDP) has been a focal point, with numerous studies exploring the Environmental Kuznets Curve (EKC) hypothesis, which postulates an inverted U-shaped relationship between per capita income and environmental degradation \citep{marjanovic2016prediction}. Industrial activity, another crucial economic indicator, has been linked with higher carbon emission due to the energy-intensive nature of manufacturing processes \citep{zhang2020impacts}. Energy consumption, particularly the reliance on fossil fuels, has been consistently identified as a primary indicator of carbon emission. The shift towards renewable energy sources is reflected in the reduced emission intensity in various predictive studies \citep{heydari2019renewable}. Technological innovation, often measured through indicators such as energy efficiency improvements and the adoption of cleaner technologies, has also been shown to have a mitigating effect on emission \citep{yu2019impact}. Societal behaviors and policy indicators, including urbanization rates, population density, and environmental regulations, have emerged as significant in the literature. These factors influence energy demand and consumption patterns, thereby affecting carbon emission \citep{ye2024industrial}. Furthermore, the implementation of carbon pricing mechanisms, such as carbon taxes and emission trading schemes, has been analyzed as an indicator for incentivizing emission reductions \citep{zhao2021multi}. The identification of these indicators is critical for policymakers to target the most influential factors that can lead to substantial emission reductions. The predictive models that incorporate these indicators provide a basis for scenario analysis, enabling decision-makers to forecast the outcomes of various policy interventions and to strategize effectively for climate change mitigation \citep{ameyaw2020west}. Hence, the identification and analysis of indicators in carbon emission prediction are essential components of climate change research. They provide the quantitative underpinning for understanding emission dynamics and crafting strategies to meet global climate goals. The ongoing research in this domain continues to refine these indicators and improve the predictive power of models, thereby contributing to more informed and effective environmental policymaking.

In exploring CO2 emission predictions, a range of methodologies has been applied, each revealing distinct advantages and inherent limitations. In one study, author utilizes Artificial Neural Networks (ANN) and Vector Autoregression (VAR) for forecasting emission in the USA, shedding light on the impact of economic growth and energy consumption on CO2 levels. However, the study concedes its inability to accurately predict output spike forms, suggesting potential inaccuracies during abrupt economic or environmental shifts \citep{mutascu2022co2}. In another study, authors adopted System Dynamics (SD) with STELLA’s simulation for city-level CO2 forecasting in Japan, achieving high accuracy. Yet, their model's assumption of a relatively constant urban structure might not fully capture the dynamic and evolving urban landscapes \citep{lee2021forecasting}.

Regression analyses have been employed to investigate GHG emission trends across OECD countries, noting the role of energy prices and economic output. They acknowledge the possibility of specification errors and the difficulty in capturing the complex drivers of emission across various national contexts \citep{calbick2014differences}. Another study investigated the influential factors of carbon emission intensity using symbolic regression, pinpointing GDP, population, and urbanization as critical factors. Nonetheless, the study's emphasis on factor occurrence over precise impact restricts a deeper understanding of causality and influence magnitude on CO2 emission \citep{pan2019influential}. These efforts collectively highlight the intricate challenges in predicting CO2 emission, from specific model limitations to broader data availability issues and the dynamic interplay of socio-economic and environmental factors.

Regarding CO2 emission clustering, diverse methodologies showcase their unique constraints. One study employs DEA window analysis and Moran’s I statistic for spatial clustering in China's provinces \citep{zhang2016co2}, while another leverages Symbolic Regression for Regression Discontinuity (SRRD) alongside the Apriori algorithm, clustering countries by their CO2 emission and economic growth models, thus innovatively identifying underlying models \citep{li2019cluster}. Further, Structural Decomposition Analysis (SDA) is used to dissect factors driving global GHG emission growth, providing insights at both country and industry levels \citep{liu2019causes}. Despite these advancements in clustering methodologies, the integration of both prediction and clustering approaches remains unexplored, presenting a gap in the literature that merits further investigation for a more comprehensive understanding of CO2 emission dynamics.

Table \ref{tab:research_summary} provides a comprehensive overview of various research studies focused on CO2 emission analysis, categorizing them based on the methodologies employed, their key findings, and the application of machine learning techniques. Specifically, it delineates whether each study incorporated Supervised Learning (SL), utilized Panel Data (PD), considered Unsupervised Learning (UL), and analyzed Multiple Countries (MC). This classification helps in identifying methodological gaps and trends within the existing literature. Upon analyzing Table 1, it becomes evident that while numerous studies have contributed to the field of CO2 emission prediction and clustering, there is a distinct lack of research that simultaneously integrates SL, PD, UL, and MC analysis within a single framework. Recognizing this gap, our research aims to address it by combining these elements to provide a more holistic and comprehensive analysis of CO2 emission. This integrative approach is designed to leverage the strengths of both supervised and unsupervised learning, while the use of panel data from multiple countries will enhance the generalizability and applicability of our findings across different geographical and economic contexts.

\begin{sidewaystable}
\caption{Summary of Methods and Findings in Carbon Emission Prediction Research}\label{tab:research_summary}
\begin{tabular*}{\textheight}{@{\extracolsep\fill}|p{0.07\textwidth}|p{0.18\textwidth}|p{0.25\textwidth}|p{0.13\textwidth}|p{0.25\textwidth}|}
\toprule
\textbf{Reference} & \textbf{Method} & \textbf{Finding} & \textbf{SL/PD/UL/MC} & \textbf{Features} \\
\midrule
\cite{mutascu2022co2} & Artificial Neural Network and Vector Autoregressive Estimator & Importance of renewable energy consumption on CO2 emission & Y/N/N/N & Economic growth, net trade, renewable energy types \\
\cite{zhao2018forecasting} & Mixed Data Sampling Regression and Back Propagation Neural Network & Accuracy improvement using hybrid forecasting model & Y/N/N/N & Quarterly economic growth data to predict annual CO2 emission \\
\cite{bennedsen2021modeling} & Structural Augmented Dynamic Factor Model & Model effectively forecasts and nowcasts U.S. CO2 emission & Y/N/N/N & Macroeconomic variables, specifically industrial production indices and residential utilities \\
\cite{wang2020decomposition} & Extended Logarithmic Mean Divisia Index Decomposition & Main factor increasing US emission is the scale effect, while the technology effect significantly reduces emission & Y/N/N/N & Income, population, energy intensity, emission coefficient, economic structure, and energy consumption structure \\
\cite{wang2019regional} & Gray Wolf Optimizer combined with a Support Vector Machine & New model effectively predicts carbon emission, showing better performance compared to traditional models & Y/N/N/N & Economic growth, population, energy consumption, industrial structure, urbanization rate, and technological advancement \\
\cite{tian2022research} & STIRPAT Model with Ridge Regression and Scenario Analysis & Differential peak attainment of carbon emission among 16 cities & Y/Y/N/N & 16 socio-economic factors including GDP, population, industry output values, and energy consumption \\
\cite{huber2021carbon} & Linear Regression for marginal emission factors, and a Multilayer Perceptron & Smart charging reduces CO2 emission by 1\% to 10\%, by shifting to times with lower marginal factors & Y/N/N/N & Historical load data, temperature data, calendar information, and possibly other not specified variables for forecasting \\
\cite{ashina2012roadmap} & AIM/Backcasting Model & Achieving an 80\% reduction in CO2 emission by 2050 is feasible with early actions & Y/N/N/N & Consumer preferences, technological options including current and future best available technologies, socio-economic conditions \\
\cite{lee2021forecasting} & System Dynamics using STELLA’s simulation engine & High forecasting accuracy with MAPEs less than 20\% for all activities & Y/N/N/N & Socio-economic data, residential statistics, service and industrial output, transportation data \\
\cite{dong2018probability} & Back propagation Neural Networks & CO2 emission are predicted to increase by 26.5–36.5\% by 2030 & Y/N/N/Y & Economic growth, energy consumption, share of renewable energy \\
\cite{alotaibi2021using} & Quantile Regression & Negative marginal relationship between CO2 emission and both socioeconomic indicators & Y/Y/N/Y & GDP per capita, fossil fuel consumption, urbanization, trade openness, population density, LPI, and HDI \\

\bottomrule
\end{tabular*}
\end{sidewaystable}

\begin{sidewaystable}
\caption*{\textbf{Table} \ref{tab:research_summary}: Summary of Methods and Findings in Carbon Emission Prediction Research (Continued)} 
\centering
\begin{tabular*}{\textheight}{@{\extracolsep\fill}|p{0.07\textwidth}|p{0.18\textwidth}|p{0.25\textwidth}|p{0.13\textwidth}|p{0.25\textwidth}|}
\toprule
\textbf{Reference} & \textbf{Method} & \textbf{Finding} & \textbf{SL/PD/UL/MC} & \textbf{Features} \\
\midrule
\cite{zhang2023novel} & Sparrow Search Algorithm, Fractional Accumulation Grey Model, and Support Vector Regression & Significantly improves prediction accuracy of carbon emission & Y/Y/N/Y & Economic, social, technological, energy, and trade factors \\
\cite{shah2023role} & Slack-Based Measure Data Envelopment Analysis and Malmquist-Luenberger Index & Renewable energy consumption increases average energy efficiency but decreases energy productivity & Y/Y/N/Y & Labor, gross fixed capital formation, non-renewable energy consumption, renewable energy consumption, GDP, and CO2 emission \\
\cite{calbick2014differences} & Series of Regression analyses & Energy prices, economic output, and environmental governance are key in explaining GHG emission variations & Y/Y/N/Y & Climate, population pressure, economic output, technological development, industrial structure, energy prices, environmental governance, pollution abatement and control expenditures, environmental pricing \\
\cite{pan2019influential} & Symbolic Regression & Influential factors vary by country, with GDP being the most frequent & Y/Y/N/Y & GDP, total population, industrialization, urbanization, technological innovation, and foreign direct investment \\
\cite{wang2021impacts} & Dynamic Panel Autoregressive Distribution Lag model & Urbanization has a negative impact on CO2 emission per capita, total CO2 emission, and CO2 emission intensity in OECD countries & Y/Y/N/Y & Urbanization rate, GDP per capita, energy intensity, industry, transportation, residential sectors, population \\
\cite{zhang2016co2} & The DEA window analysis approach and Moran’s Index & The northwest region has the highest CO2 emission reduction potential in the industrial sector & N/Y/Y/N & Labor, capital, energy consumption, value-added of industrial enterprises, and CO2 emission \\
\cite{gallo2018clustering} & Statistical Analysis and Clustering & Kyoto Protocol has not considered its effects on agricultural and forestry & N/Y/Y/Y & Greenhouse gas emission data from the agricultural and forestry sectors \\
\cite{lamb2022countries} & Linear Model \& Natural Logarithm & 24 countries managed to reduce both CO2 and GHG emission & N/N/Y/Y & CO2 and GHG emission \\
\cite{anwar2020impact} & Panel data-fixed effect model, incorporating time trends & Urbanization, economic growth, and trade openness significantly determine CO2 emission & Y/Y/N/Y & Urbanization, GDP, trade openness \\
\cite{liu2019causes} & Structural Decomposition Analysis & Rapid global economic growth is a dominating driving force & N/N/Y/Y & Industry energy use, CO2 emission and emission to air in the world \\
Our Proposed Approach & Five statistical models, SARIMAX, and DTW & Identifying important factors, 3 years CO2 Emission prediction into the future, clustering 20 countries in 3 groups & Y/Y/Y/Y & Mentioned in Table \ref{tab:indicators} \\
\bottomrule
\end{tabular*}
\end{sidewaystable}

\section{Data Understanding}\label{sec3}

For the development of a global CO2 emission prediction model, we have meticulously sourced our data from the World Development Indicators  database, a comprehensive repository maintained by the World Bank \citep{worldbank2020indicators}. This dataset provides an extensive range of economic, social, and environmental indicators, which have been instrumental in understanding and forecasting global CO2 emission trends. Table \ref{tab:indicators} summarizes the information about the features and their importance in the modeling process.

\begin{figure}[!htb]
\centering
\includegraphics[width=0.95\linewidth]{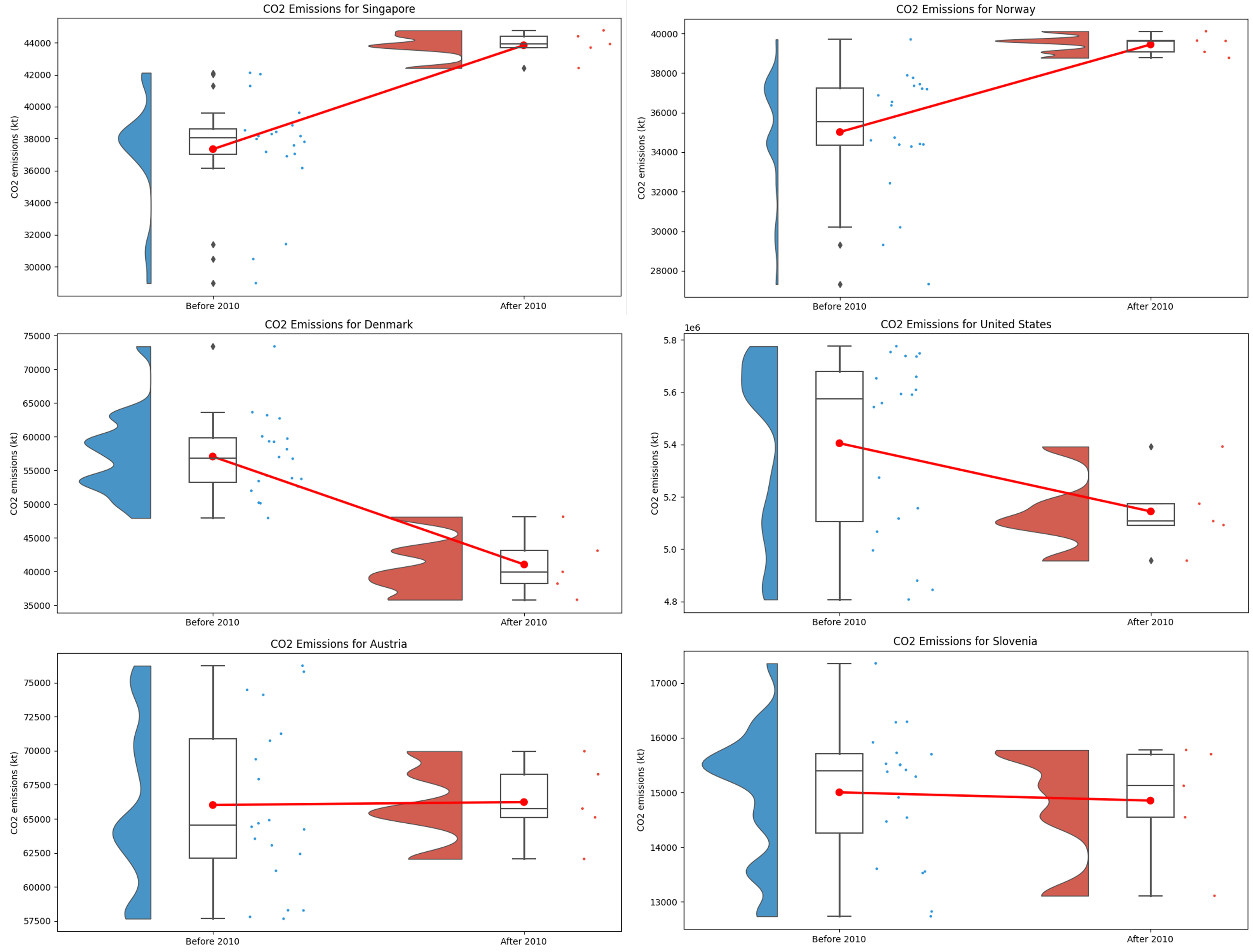}
\caption{Different carbon emission patterns over years in countries}
\label{fig:myimage3}
\end{figure}

We have attempted to visualize the data to gain a deeper understanding of the underlying patterns and correlations within the data to assist the development of the model. It is possible to identify trends, and uncover underlying structures by employing a variety of visualization techniques. Firstly, we examine the trend in CO2 emission among various countries. The available data has been compared from two distinct periods: 20 years preceding 2010 and 5 years following 2010. Understanding emission patterns over time and identifying potential shifts in environmental policy or industrial practices are crucial for understanding how emission patterns have evolved. Based on Figure~\ref{fig:myimage3}, countries exhibit three distinct emission trends. The first group, exemplified by Singapore and Norway, exhibits an upward trend in CO2 emission following 2010. As a result of this increase, emission may have increased either due to an increase in industrial activity or energy consumption. In contrast, the second group, including countries such as the United States and Denmark, exhibits a decrease in emission following 2010. As a result of successful environmental policies, the shift towards renewable energy sources, or improvements in energy efficiency, this reduction may occur. Additionally, for countries such as Slovenia and Austria, our analysis revealed a consistent trend in CO2 emission without any significant decrease or reduction observed over the two periods.

\begin{figure}[!htb]
\centering
\includegraphics[width=0.95\linewidth]{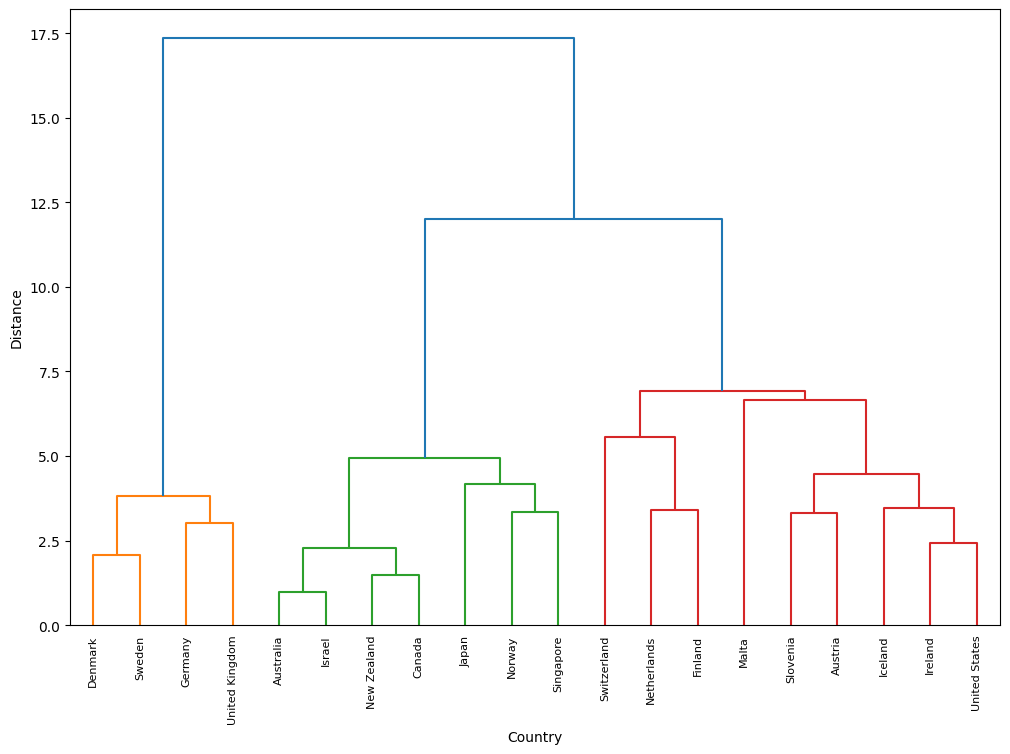}
\caption{The dendrogram visualization of grouped countries by their normalized CO2 emission}
\label{fig:myimage4}
\end{figure}

To get an overall idea about the pattern of CO2 emission, it is necessary to categorize countries based on their long-term emission trends. Analyzing trends in CO2 emission provides initial insights. However, the historical emission patterns of each country should be examined. As shown in Figure~\ref{fig:myimage4}, a hierarchical clustering approach coupled with Ward's method \citep{murtagh2014ward} demonstrates a dendrogram that provides a systematic method for grouping countries by their normalized CO2 emission. The Ward method ensures that all the countries grouped together exhibit similar emission dynamics by minimizing the total within-cluster variance. Distances between clusters are indicated by the height of the merge line, with higher heights representing greater distances. By acknowledging that countries can be divided into three distinct groups, we can create a clustering model tailored to classify the data into these three separate categories.

\begin{sidewaystable}
\caption{Indicators of Environmental Impact and Energy Use}\label{tab:indicators}
\begin{tabular*}{\textheight}{@{\extracolsep\fill}|p{0.175\textwidth}|p{0.425\textwidth}|p{0.3\textwidth}|}
\toprule
\textbf{Indicator} & \textbf{Definition} & \textbf{Significance} \\
\midrule
CO2 emission (kt) (Co) & This indicates the emissions coming from the burning of fossil fuels and the manufacture of cement. Includes carbon dioxide produced during consumption of solid, liquid, and gas fuels and gas flaring. & Direct indicator of a country's contribution to global CO2 levels and climate change. \\
CO2 emission (metric tons per capita) & Carbon dioxide emissions in metric tons per capita as defined above. & Reflects the average CO2 footprint of individuals, useful for comparing among populations. \\
Energy use (kg of oil equivalent per capita) (EU) & Energy use calculated as primary energy before transformation to other end-use fuels, equal to indigenous production plus imports and stock changes, minus exports and fuels supplied to ships and aircraft engaged in international transport. & Higher per capita energy use typically correlates with higher CO2 emission. \\
Renewable energy consumption (\% of total final energy consumption) (RE) & The share of renewable energy in total final energy consumption. & The greater the reliance on renewables, the lower the CO2 emission from energy consumption. \\
Electricity production from oil, gas, and coal sources (\% of total) (EP) & Electricity production sources are the fuels used to generate electricity. Includes crude oil, petroleum products, natural gas, coal, and brown coal, both primary and derived fuels like coke. & Indicates the dependency on fossil fuels for electricity, which is directly related to CO2 emissions and environmental impact. \\
Forest area (sq. km) (F) & Land under natural or planted stands of trees of at least 5 meters in situ, excluding trees in agricultural systems and urban areas. & Forests are carbon sinks; larger forest areas can mitigate CO2 emission by absorbing more CO2. \\
Forest area (\% of land area) (Fa) & Proportion of land covered by forests. & Higher percentages indicate greater potential for natural CO2 absorption and climate regulation. \\
Electric power consumption (kWh per capita) (EPC) & Measures the production of power plants less transmission, distribution, and transformation losses and own use by heat and power plants. & High per capita consumption often leads to higher CO2 emission if electricity is produced from fossil fuels. \\
GDP per unit of energy use (PPP \$ per kg of oil equivalent) (G) & PPP GDP per kilogram of oil equivalent of energy use. An international dollar has the same purchasing power over GDP as a U.S. dollar has in the United States. & Efficient use of energy for economic output can suggest a lower CO2 emission intensity. \\
Total greenhouse gas emission (kt of CO2 equivalent) (TG) & Includes all anthropogenic CH4 sources, N2O sources, and F-gases (HFCs, PFCs, and SF6), but excludes short-cycle biomass burning. & Provides a broader picture of a country's impact on warming, including but not limited to CO2 emission. \\
\bottomrule
\end{tabular*}
\end{sidewaystable}

\begin{figure}[!htb]
\centering
\includegraphics[width=0.75\linewidth]{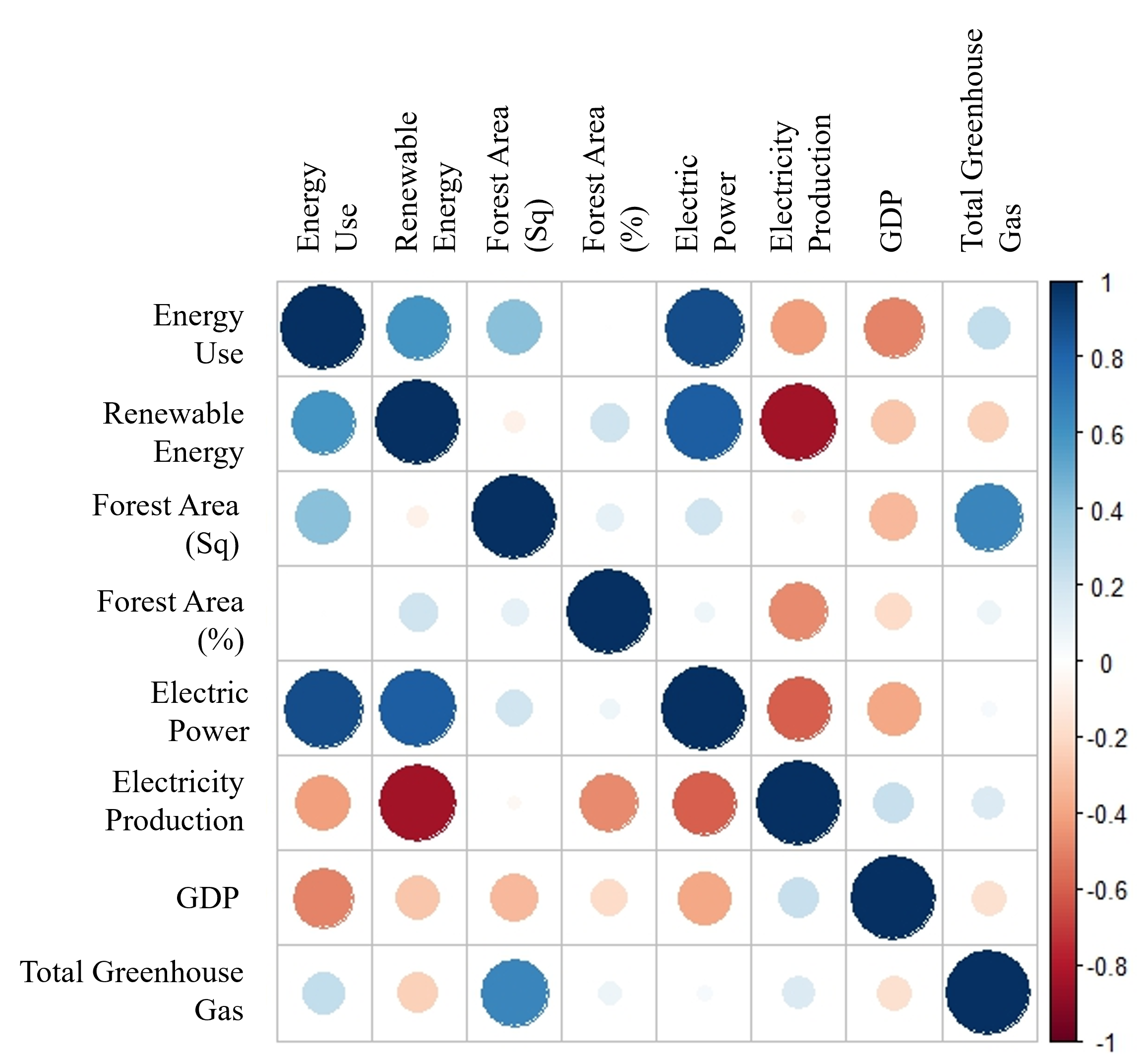}
\caption{The heatmap visualization of the features}
\label{fig:myimage5}
\end{figure}

\begin{figure}[!htb]
\centering
\includegraphics[width=0.95\linewidth]{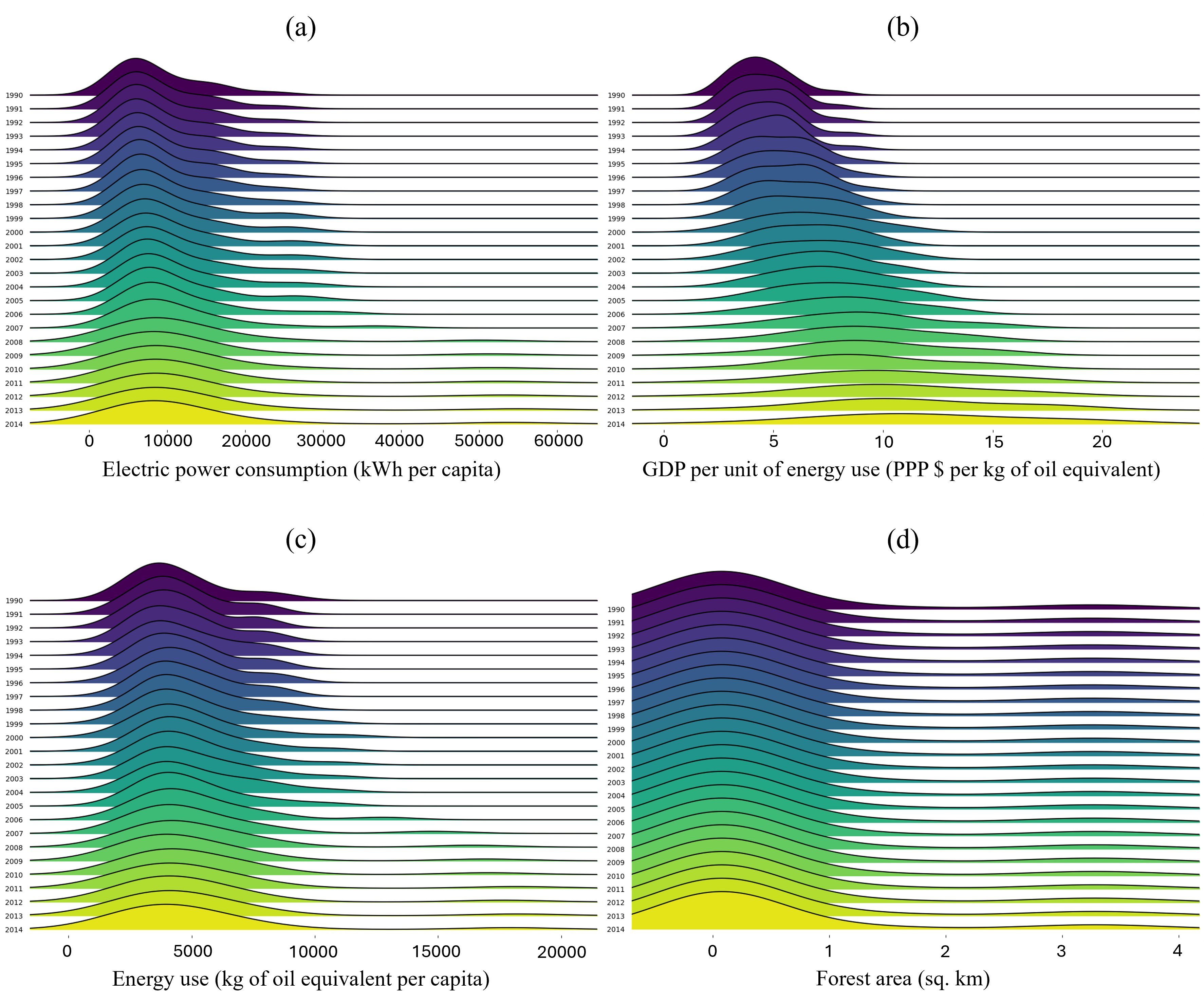}
\caption{The ridge plot of the four indicators showing the distribution and evolution of them over time}
\label{fig:myimage6}
\end{figure}

To analyze the features and their correlations, Figure~\ref{fig:myimage5} is presented. Figure~\ref{fig:myimage5} provides a correlation matrix, the size and color of the circles indicate the strength and direction of the correlation between the variables. In this plot, the larger circles represent stronger correlations, whereas blue represents positive correlations and red indicates negative correlations. It is noteworthy that `Renewable energy consumption' and `Electric power consumption' exhibit a highly positive correlation (quantified at 0.82). Also, there is a notable positive correlation between `Electric power consumption' and `Energy use,' evidenced by a correlation coefficient of 0.88. Conversely, a strong negative correlation of -0.83 is observed between `Electricity production from oil, gas, and coal sources' and `Renewable energy consumption.' This inverse relationship suggests that as one variable increases, the other tends to decrease, possibly reflecting a shift towards renewable energy sources or improvements in production efficiency. These correlations have a considerable impact on our modeling approach. When both features, highly correlated, emerge as significant predictors, it is more suitable to use only one of them to avoid multicollinearity. Multicollinearity can lead to skewed predictions and hinder the interpretability of the model, thus careful selection of predictors is crucial to maintain the accuracy and reliability of our analysis. To further understand the distribution and evolution of the indicators over time, Figure~\ref{fig:myimage6} is presented.

\begin{figure}[!htb]
\centering
\includegraphics[width=1.0\linewidth]{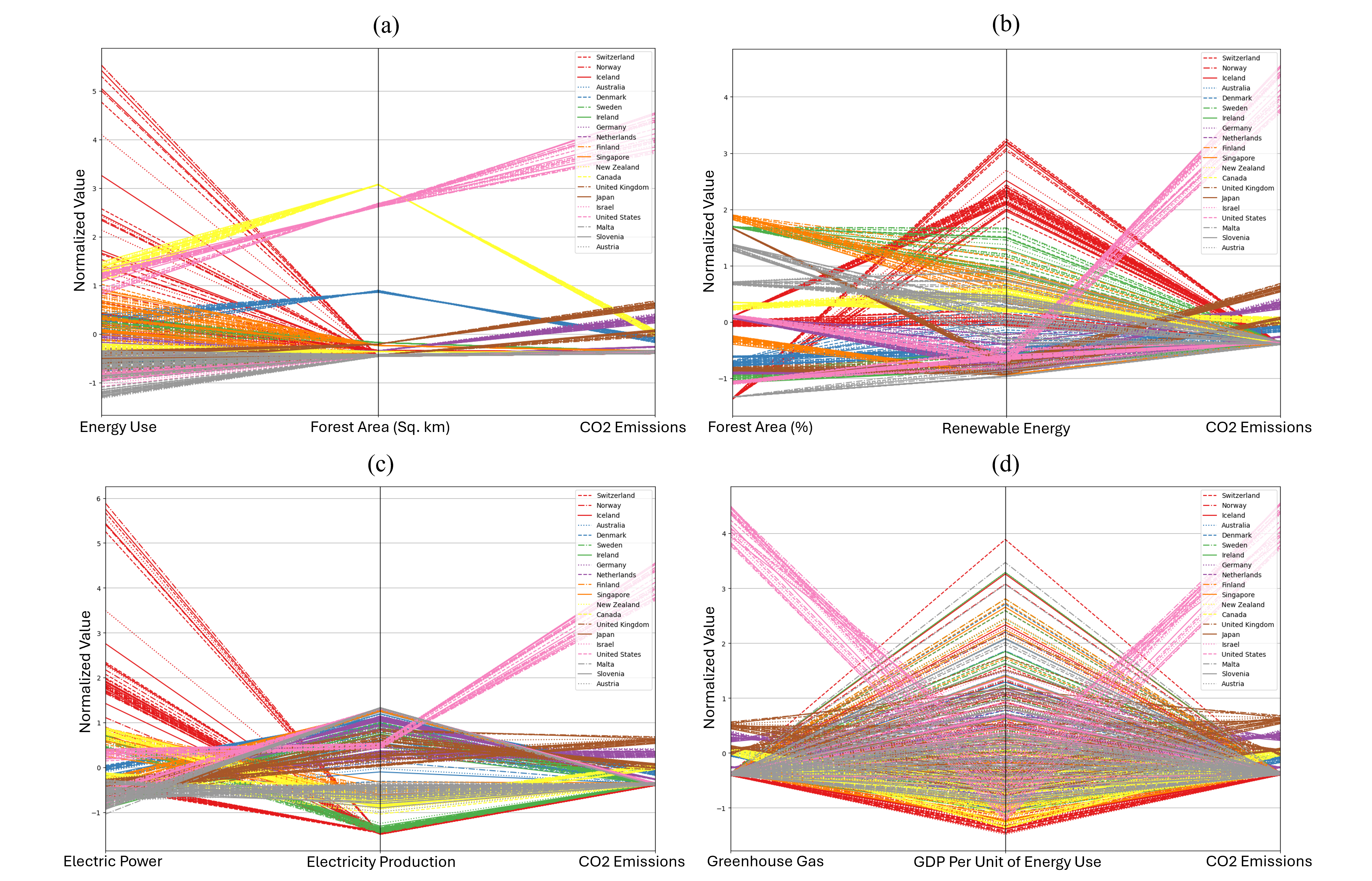}
\caption{The parallel coordinates plots of the features to provide a multi-dimensional perspective }
\label{fig:myimage7}
\end{figure}

Using ridge plots in Figure~\ref{fig:myimage6}, we present insight into four characteristics of our panel dataset based on their distributional characteristics. The Figure~\ref{fig:myimage6}(a) illustrates the consumption of electricity over time. The layer represents a specific year, and the width indicates the density of countries consuming specific amounts of electric power. More countries are consuming larger quantities of electricity in the last years, as indicated by the widening of the plot. Figure~\ref{fig:myimage6}(b) demonstrates that the layers in the `GDP per unit of energy use' ridge plot progressively widened and shifted rightward from 1990 to 2014. This trend indicates an increasing variability among countries in their efficiency of energy use for GDP production during that period. Figure~\ref{fig:myimage6}(c) presents considerable variation in the height of layers for different years, depicting energy use trends. The horizontal spread of these layers signifies a general increase in energy consumption, though at varying rates among different countries. This variation underscores the diverse energy consumption trajectories globally. In Figure~\ref{fig:myimage6}(d), the depiction of changes in forest area over the past 25 years is marked by distinct annual layers. Notably, the peak of these layers maintains a relatively stable position, without significant shifts to the right or left. This stability suggests that there hasn't been a dramatic change in forest areas globally over the observed period. Additionally, the height of the layers does not show marked fluctuations, indicating a relative steadiness in forest coverage year over year. Next, parallel coordinates plots are demonstrated to provide a multi-dimensional perspective on how various factors relate to CO2 emission in Figure~\ref{fig:myimage7}. Across these visualizations, each line corresponds to a country, while the position along each axis represents a country's normalized value for the features.

According to Figure~\ref{fig:myimage7}(a), higher levels of energy consumption do not necessarily lead to higher CO2 emission across countries, suggesting the importance of other factors. Similarly, Figure~\ref{fig:myimage7}(b) does not indicate a clear correlation between `Forest area (\%)', `Renewable energy', and CO2 emission, meaning that there exist other significant contributions such as industrial activity levels or environmental regulations, as evidenced by intersecting paths between these axes. Based on the Figure~\ref{fig:myimage7}(c), there is no direct relationship between `Electric power', `Electricity production from oil, gas, and coal sources', and CO2 emission either. However, in the Figure~\ref{fig:myimage7}(d), the pattern is more predictable. In this plot, a lower `GDP per unit of energy use' correlates with higher greenhouse gas emission, leading to higher CO2 emission.
These visualizations, while insightful, highlight the challenges in identifying the most influential factors for predicting CO2 emission. Therefore, a comprehensive feature selection process, coupled with advanced modeling techniques for panel data, is essential to accurately pinpoint the key drivers of CO2 emission. This approach is vital for developing effective machine learning models that can navigate the complex factors influencing CO2 emission.

\section{Methodology}\label{sec4}

In this study, we have employed a two-phase analytical approach to analyze CO2 emission. Initially, five statistical models including OLS, fixed effects, and random effects are used to identify key features affecting CO2 emission. Subsequently, we implement both supervised (SARIMAX) and unsupervised (DTW) learning models for a thorough examination of emission trends.

Our methodological advancements address limitations of traditional linear regression models in handling panel dataset. By using models that account for unobserved heterogeneity across countries and time, we have achieved more robust inferences, crucial given the diversity and dynamic nature of high-HDI countries' emission. SARIMAX’s (which stands for Seasonal AutoRegressive Integrated Moving Average with eXogenous variables) ability to integrate seasonal patterns and external variables offers a nuanced understanding of emission, while DTW’s refined clustering technique provides insights into countries’ emission trajectories, highlighting convergences and divergences in their patterns over time. This comprehensive methodology facilitates a deeper understanding of the factors driving CO2 emission, enhancing the potential for effective policy formulation.

\subsection{Terminology}\label{subsec4}

\subsubsection{Pooled OLS Model}\label{subsubsec4}

Pooled Ordinary Least Squares (Pooled OLS) is a prevalent econometric technique for analyzing panel data, which involves multiple observations over time for a number of cross-sectional units. This method is particularly useful for datasets where the focus is on understanding the collective impact of variables across different entities, such as countries or firms, over time. Pooled OLS treats the panel data as a large cross-sectional dataset, essentially aggregating all time periods and entities, which simplifies the analysis but also imposes certain limitations. The core idea of Pooled OLS is to estimate a single regression model for the entire panel dataset. In Pooled OLS, the regression equation is formulated as:

\begin{equation}
\label{eq1}
Y_{it} = \beta_{0} + \beta_{1}X_{it} + \epsilon_{it}
\end{equation}

In the above equation, $Y_{it}$ is the dependent variable for the $i$-th cross-sectional unit (e.g., country) at time $t$, $X_{it}$ represents independent variables, $\beta_0$ and $\beta_1$ are coefficients, and $\epsilon_{it}$ is the error term. This model assumes that the slope coefficients ($\beta$) are constant across all cross-sectional units and over time. The estimation of the model parameters is typically achieved through OLS method, which involves minimizing the sum of squared differences between the observed values and the values predicted by the model. One of the primary limitations of Pooled OLS is its assumption of homogeneity, implying that the slope coefficients are constant across all entities and over time. This assumption often leads to biased estimates in the presence of unobserved heterogeneity among cross-sectional units. Moreover, Pooled OLS assumes no autocorrelation and homoscedasticity in error terms, which is often violated in panel data settings. Recent studies emphasize the need for careful application of Pooled OLS, especially in the presence of cross-sectional dependence and heterogeneity.

\subsubsection{Random Effects Model}

The Random Effects (RE) model stands as a pivotal approach in the realm of panel data analysis, particularly when addressing data that comprises multiple observations across various entities over time. Distinguished from the Pooled OLS approach, the RE model acknowledges the potential heterogeneity across cross-sectional units (such as countries, companies, or individuals) by incorporating a unique error component for each entity. This model is especially pertinent when the unobserved entity-specific effects are assumed to be uncorrelated with the explanatory variables, a condition that makes it distinct from the Fixed Effects (FE) model. The RE model is predicated on the notion that while all entities in the panel share common behavior captured by the overall intercept and slope coefficients, there is also an idiosyncratic, entity-specific random component. The model is typically represented as:

\begin{equation}
\label{eq2}
Y_{it} = \beta_{0} + \beta_{1}X_{it} + u_{i} + \epsilon_{it}
\end{equation}

Here, $Y_{it}$ denotes the dependent variable for entity $i$ at time $t$, $X_{it}$ represents the independent variables, $\beta$s are the coefficients to be estimated, $u_{i}$ is the unobserved random effect unique to each entity, and $\epsilon_{it}$ is the idiosyncratic error term. The estimation of the RE model usually involves Generalized Least Squares (GLS), a method that efficiently deals with the panel-specific variance component $u_{i}$. The critical limitation of the RE model lies in its assumption of no correlation between the entity-specific effects and the independent variables. If this assumption is violated, the RE model may yield biased and inconsistent estimates. Recent methodological advancements in panel data analysis have focused on developing more robust random effects models that can handle situations where traditional RE assumptions might not hold.

\subsubsection{Fixed Effects Model}

The FE model is a cornerstone in panel data analysis, particularly valuable when dealing with data that exhibits variability across entities and over time. This model is particularly adept at analyzing effects within entities (such as countries, firms, or individuals) by controlling for entity-specific characteristics. The FE model is distinguished by its ability to account for unobserved heterogeneity when this heterogeneity is correlated with the independent variables in the model, a key point that differentiates it from the RE model.
The FE model operates under the premise that there are unique attributes inherent to each entity that may influence the dependent variable. These attributes, while potentially correlated with the independent variables, are assumed to be constant over time. The model can be represented as:

\begin{equation}
\label{eq3}
Y_{it} = \alpha_{i} + \beta X_{it} + \epsilon_{it}
\end{equation}

Here, $Y_{it}$ represents the dependent variable for entity $i$ at time $t$, $X_{it}$ are the independent variables, $\beta$ represents the coefficients associated with each independent variable, which are to be estimated, $\alpha_{i}$ symbolizes the unobserved individual-specific effect, and $\epsilon_{it}$ is the idiosyncratic error term. FE models are typically estimated using Least Squares Dummy Variable (LSDV) approach or the within transformation. The LSDV approach involves including a dummy variable for each entity, whereas the within transformation, more commonly used due to its efficiency, removes the entity-specific effects by demeaning the variables by their means within each entity. This transformation focuses on the deviations from each entity’s mean, eliminating the entity-specific effects and isolating the impact of the independent variables.

\subsubsection{Dynamic Time Warping}

In this research, Dynamic Time Warping (DTW) is employed as a key methodology to cluster countries based on the trends in their CO\(_2\) emission. This approach is particularly suited for analyzing panel datasets where temporal patterns in CO\(_2\) emission across countries may not align perfectly in time. DTW's ability to measure similarities between temporal sequences with varying lengths and speeds makes it an ideal tool for grouping countries with similar emission trends, despite discrepancies in timing or rate of change. DTW is used to compare the temporal sequences of CO\(_2\) emission between different countries. Each country's emission data forms a time series, \(X=\{x_1,x_2,\ldots,x_n\}\), and DTW measures the similarity of these time series by aligning them in a manner that minimizes the cumulative distance. This is achieved by constructing a matrix where each element represents the distance between two points from two different time series (i.e., two different countries' CO\(_2\) emission levels at different times).

\subsubsection{SARIMAX}

The ARIMA (Autoregressive Integrated Moving Average) model is a cornerstone in time series forecasting, particularly suited for data showing trends and non-seasonal patterns. It is expressed as ARIMA\((p,d,q)\), where \(p\) is the order of the autoregressive part, \(d\) the degree of differencing, and \(q\) the order of the moving average part. The model can be represented as:

\begin{equation}
\label{eq4}
(1-\sum_{i=1}^{p} \phi_i L^i) (1-L)^d Y_t = (1+\sum_{i=1}^{q} \theta_i L^i) \epsilon_t
\end{equation}

Here, $Y_t$ is the time series, $L$ is the lag operator, $\phi_i$ are the parameters of the autoregressive part, $\theta_i$ of the moving average part, and $\epsilon_t$ is white noise error. While ARIMA is powerful, it has limitations in handling seasonal fluctuations and external influences, which are common in many time series data, including CO2 emission. This leads to the evolution into SARIMAX (Seasonal Autoregressive Integrated Moving Average with eXogenous variables), an extension of ARIMA that incorporates both seasonality and exogenous factors. SARIMAX is denoted as SARIMAX$(p,d,q)(P,D,Q,S)$, where $P, D, Q$ represent the seasonal elements of the AR, differencing, and MA components, respectively, and $S$ is the periodicity of the seasons.

\begin{equation}
\label{eq5}
\begin{split}
&(1-\sum_{i=1}^{p} \phi_i L^i)(1-\sum_{i=1}^{P} \Phi_i L^{iS}) \times (1-L)^d (1-L^S)^D Y_t = \\
&(1+\sum_{i=1}^{q} \theta_i L^i)(1+\sum_{i=1}^{Q} \Theta_i L^{iS}) \epsilon_t + \beta X_t
\end{split}
\end{equation}

Here, $\Phi_i$ and $\Theta_i$ are the seasonal AR and MA parameters, $D$ is the order of seasonal differencing, $\beta$ represents the coefficients of the exogenous variables $X_t$, and $S$ denotes the seasonality period. SARIMAX offers a substantial advancement over ARIMA in time series forecasting, particularly for complex datasets like CO2 emission, due to its enhanced capabilities. It adeptly handles seasonal trends through seasonal differencing and the incorporation of seasonal autoregressive and moving average components, a critical aspect for data exhibiting seasonal variations. Additionally, SARIMAX's ability to include exogenous variables allows for the integration of external factors, enhancing the robustness and comprehensiveness of the model.

\subsubsection{MAE}

MAE measures the average magnitude of errors in a set of predictions, without considering their direction. It’s a linear score, which means all the individual differences are weighted equally. It is calculated as:

\begin{equation}
\label{eq6}
\text{MAE} = \frac{1}{N} \sum_{i=1}^{N} |y_i - \hat{y}_i|
\end{equation}

In the above equation, $N$ is the number of observations, $y_i$ is the actual value of the observation, and $\hat{y}_i$ is the predicted value.

\subsubsection{RMSE}

RMSE is a standard way to measure the error of a model in predicting quantitative data. It is particularly sensitive to large errors due to its squaring of the residuals. The RMSE is calculated as:

\begin{equation}
\label{eq7}
\text{RMSE} = \sqrt{\frac{1}{N} \sum_{i=1}^{N} (y_i - \hat{y}_i)^2}
\end{equation}

In the above equation, $N$ is the number of observations, $y_i$ is the actual value of the observation, and $\hat{y}_i$ is the predicted value.

\subsubsection{NRMSE}

NRMSE is a normalized version of RMSE that facilitates the comparison between models with different scales. It is defined as:

\begin{equation}
\label{eq8}
\text{NRMSE} = \frac{\text{RMSE}}{y_{\max} - y_{\min}}
\end{equation}

Here, RMSE is the root mean square error, $y_{\max}$ and $y_{\min}$ are the maximum and minimum values of the observed data, respectively. This normalization helps to bring the error metric to a comparable scale, especially useful when dealing with different datasets or units.

\subsection{The Proposed Method}

\begin{figure}[!htb]
\centering
\includegraphics[width=0.95\linewidth]{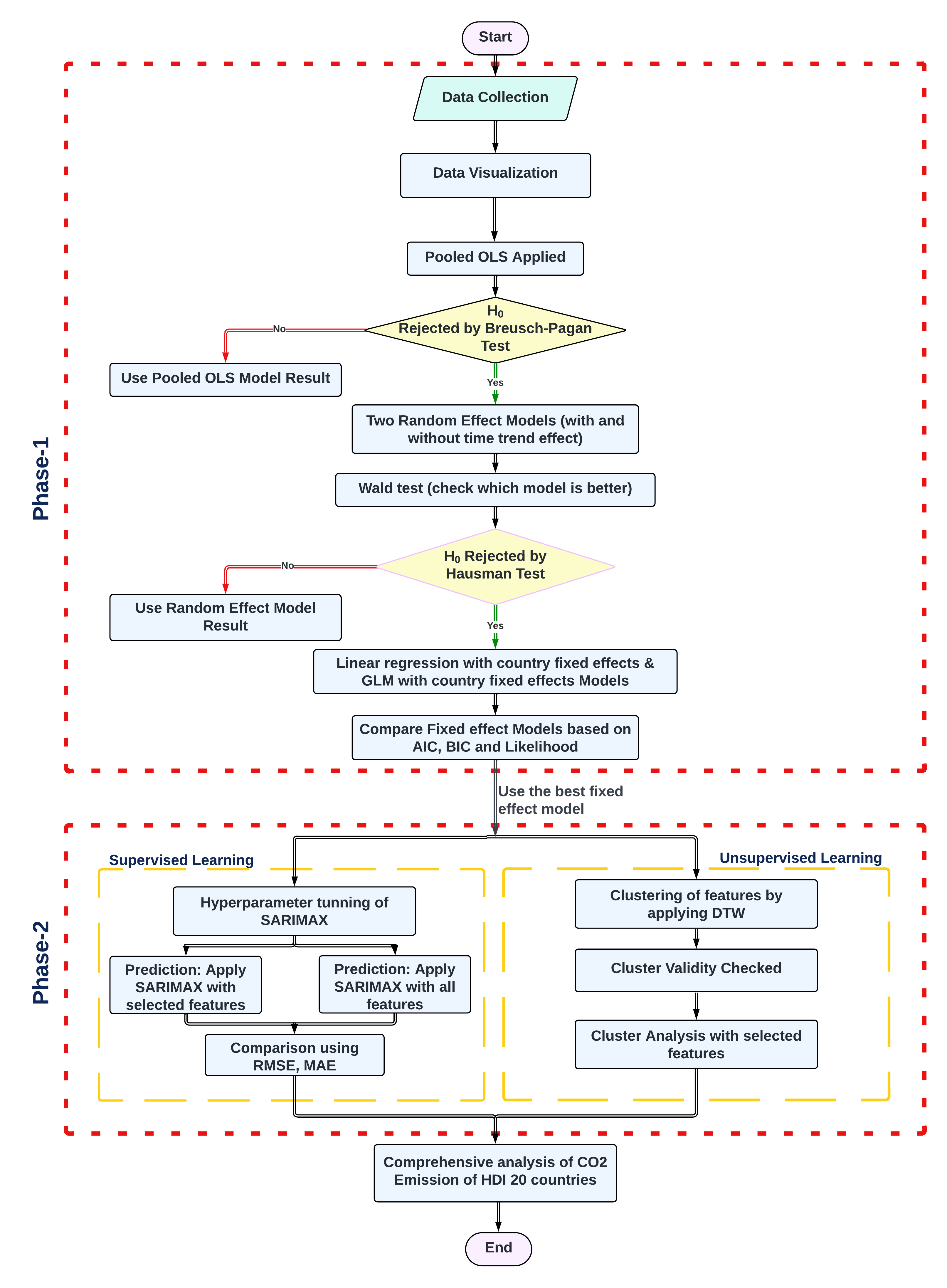}
\caption{The flow diagram of the proposed method}
\label{fig:myimage8}
\end{figure}

The proposed method presents a systematic approach to analyze CO2 emission across multiple countries over time, employing a sequence of econometric models and statistical tests. The overview of the method is shown in Figure \ref{fig:myimage8}. This starts with data collection and preparation, which is essential to ensure the integrity of the subsequent econometric analysis. The initial examination begins with a Pooled OLS model, assuming homogeneity across entities and time. Then, the Breusch-Pagan test is utilized to check whether evidence of heteroscedasticity is found or not. If this statistical test is rejected, then two different RE models are applied (with and without time trend effect). To further refine the model selection, the Wald test is applied to compare the RE models \citep{austin2018effect}. Following this, the Hausman test is conducted to find out whether fixed effect models are needed or not. Upon this test's indication, we either continue with the RE model or switch to the FE model. We then apply and compare two different fixed effect models (linear regression with country fixed effects and Generalized Linear Model (GLM) with country fixed effects) using criteria such as the AIC, BIC, and Likelihood to determine the best model in terms of fitting the data. Based on the best model found in this step, the important features affecting the CO2 emission are selected. Next, the phase two of the methodology is employed focusing on an integrated analysis using both supervised and unsupervised learning paths. It provides a thorough and robust examination of CO2 emission across countries with different Human Development Index (HDI) scores. SARIMAX and DTW are applied in this phase using the selected features from phase one. This comprehensive approach not only bolsters the analytical depth but also ensures that our results are robust and can inform impactful policy decisions within the realm of environmental climate change mitigation.

\section{Results}\label{sec5}

In this section, we present a detailed analysis of the outcomes derived from the application of outlined methodologies. Initially, we interpret the findings from the five statistical models, followed by a comparative assessment of their performance. Subsequently, leveraging the most effective model identified, we select the most relevant features that significantly influence our analysis. These features are then employed to predict CO2 emission over the most recent three-year period of the dataset.

\subsection{Phase I: Applying Specialized Models for Panel Dataset}

Our analysis comprehensively evaluates various models to ascertain their suitability for the panel dataset. We have explored five distinct models, encompassing a pooled model, two random effect models, and two fixed effect models, to understand their efficacy and limitations. The models' summaries are encapsulated in Table \ref{tab:models}, providing insights into their types, objectives, limitations, and potential biases.

\begin{table}[h]
\centering
\caption{Summary of the applied models on the panel dataset}\label{tab:models}
\begin{tabular*}{\textwidth}{@{\extracolsep\fill}l p{0.2\textwidth} p{0.2\textwidth} p{0.2\textwidth} p{0.2\textwidth}}
\toprule
\textbf{Model} & \textbf{Type} & \textbf{Objective} & \textbf{Limitation} & \textbf{Bias} \\
\midrule
A & Pooled OLS & Averages the impact of predictors & Cannot account for time-invariant characteristics & Risk of bias from correlated traits \\
B & Random-effects & Averages impacts with group variation & Assumes random effects are normally distributed & Unbiased if effects are normal \\
C & Random-effects with time trend & Considers impacts with group and time trends & Assumes effects are normally distributed & Unbiased if effects are normal \\
D & Linear regression with country fixed effects & Evaluates impacts, accounting for country specifics & Demands more data & Unbiased, controls for country traits \\
E & GLM with country fixed effects & Examines relations, incorporating fixed country traits & Assumes normal distribution of dependent variable & Unbiased, addresses country traits \\
\bottomrule
\end{tabular*}
\end{table}

As seen from Table \ref{tab:models}, Model A, the pooled model, simplifies the analysis by estimating the average effects of predictor variables without adjusting for time-invariant characteristics, making it suitable for datasets with a smaller number of entities. Model B, the random-effects model, offers a balance by assuming these characteristics are random and accounted for them as errors. Model C extends this approach by including a time trend to capture potential systematic changes over time across all groups. Model D employs linear regression with country fixed effects, enhancing the model's ability to control unobserved, time-invariant differences across countries. Lastly, Model E utilizes Generalized Linear Models (GLMs) with country fixed effects, providing a flexible framework capable of modeling a broader range of data distributions, including non-normal distributions.

\subsubsection{Pooled Ordinary Least Squares Model}

We initiate our analysis with a Pooled OLS model to assess our dataset comprehensively. The model is structured as follows:

\begin{align}
C_{oit} &= \beta_{0} + \beta_{1}RE_{it} + \beta_{2}Fa_{it} + \beta_{3}EPF_{it} + \beta_{4}G_{it} \nonumber \\
&\quad + \beta_{5}TG_{it} + \beta_{6}F_{it} + \beta_{7}EU_{it} + \beta_{8}EPC_{it} + \epsilon_{it}
\end{align}

where \(C_{oit}\) denotes the dependent variable for entity \(i\) at time \(t\), \(\beta\) values are the coefficients for the respective predictors, and \(\epsilon_{it}\) represents the error term. The regression output, summarized in Table \ref{tab:regression_summary}, reveals the significance levels of various predictors on CO2 emission. 

The analysis underscores that all variables, except `Energy use' (EU), are considered as significant predictors. The `Energy use' variable, with a p-value of 0.42, falls outside the conventional significance threshold of 0.05, suggesting it does not significantly contribute to the model. Further evaluation of the model's overall efficacy is presented in Table \ref{tab:regression_stats}.

\begin{table}[!htb]
\centering
\caption{Pooled OLS Model Results}\label{tab:regression_summary}
\begin{tabular*}{\textwidth}{@{\extracolsep\fill}lcccccc}
\toprule
Term & Estimate & Std. Error & t-value & p-value & Significance \\
\midrule
(Intercept) & -22963.0 & 14169.0 & -1.6207 & 0.1057332 & \\
EPC & 3.1854 & 0.82793 & 3.8475 & 0.0001351 & *** \\
EPF & -337.13 & 113.06 & -2.9818 & 0.0030078 & ** \\
EU & -1.6263 & 2.0231 & -0.8038 & 0.4218736 & \\
F & -0.027955 & 0.00259 & -10.7934 & $<$ 2.2e-16 & *** \\
Fa & 758.86 & 96.324 & 7.8782 & 2.153e-14 & *** \\
G & 1991.4 & 591.33 & 3.3677 & 0.0008176 & *** \\
RE & -1466.4 & 309.84 & -4.7327 & 0.000002902 & *** \\
TG & 0.85062 & 0.0016949 & 501.8764 & $<$ 2.2e-16 & *** \\
\bottomrule
\end{tabular*}
\end{table}

\begin{table}[!htb]
\centering
\caption{Pooled OLS Performance Summary}\label{tab:regression_stats}
\begin{tabular}{p{0.3\textwidth}p{0.2\textwidth}}
\toprule
\textbf{Statistic} & \textbf{Value} \\
\midrule
Total Sum of Squares & \(6.78 \times 10^{14}\) \\
Residual Sum of Squares & \(6.82 \times 10^{11}\) \\
R-Squared & 0.999 \\
Adj. R-Squared & 0.99898 \\
F-statistic & 61022.1 \\
p-value & \(< 2.22 \times 10^{-16}\) \\
\bottomrule
\end{tabular}
\end{table}

Despite the model's high R-square value and a significant F-statistic, the large sums of squares suggest potential misspecification. To address this, we applied the Breusch-Pagan test to evaluate the presence of heteroskedasticity, which could indicate a preference for a random effects model over the pooled OLS model. This test is used to test for heteroskedasticity in the residuals. When applied to panel data, it tests the null hypothesis that variances across entities (countries) are equal. If the test is significant, it suggests the presence of random effects, implying a random effects model might be more appropriate than a pooled OLS model. The hypotheses for the Breusch-Pagan test in this context are:
\begin{itemize}
    \item Null Hypothesis (\(H_0\)): There is no individual heteroskedasticity. This means that the variances across entities are equal, which implies that there are no individual-specific effects or that these effects are not correlated with other regressors.
    \item Alternative Hypothesis (\(H_1\)): There is individual heteroskedasticity. This implies the presence of individual-specific effects that might be correlated with other regressors.
\end{itemize}
The Breusch-Pagan test yielded a chi-square value of 3653.4 with a p-value \(< 2.2e-16\), compelling us to reject \(H_0\). This outcome points to heteroskedasticity and the potential suitability of a random effects model. Consequently, our next step involves a thorough investigation of random effects models to better cater to the data's characteristics.

\subsubsection{Random Effects Models}

In our exploration, we consider two distinct variants of random effects models. The initial model accommodates for unobserved heterogeneity among entities by treating these variations as random variables that are uncorrelated with the predictor variables. The subsequent model advances this approach by integrating a time trend, enabling the capture of systematic temporal changes impacting all entities. This modification aims to offer a more refined comprehension of the interplay between temporal dynamics and entity-specific variations. The formulas for these models are presented in the following. For the first model (random-effects model without time trend), we have:

\begin{equation}
\begin{split}
C_{oit} &= \beta_{0} + \beta_{1}RE_{it} + \beta_{2}Fa_{it} + \beta_{3}EPF_{it} \\
&\quad + \beta_{4}G_{it} + \beta_{5}TG_{it} + \beta_{6}F_{it} + \beta_{7}EU_{it} \\
&\quad + \beta_{8}EPC_{it} + u_{i} + \epsilon_{it}
\end{split}
\end{equation}

For the second model (random-effects model with time trend), we again have the following. In these equations, \(u_i\) denotes the unobserved random effect for entity \(i\), \(\gamma_t\) represents the fixed effect for time \(t\), and \(\epsilon_{it}\) is the idiosyncratic error term. The comparison of these models is summarized in Table \ref{tab:ModelComparison}, highlighting their fit to the data.

\begin{equation}
\begin{split}
C_{oit} &= \beta_{0} + \beta_{1}RE_{it} + \beta_{2}Fa_{it} + \beta_{3}EPF_{it} \\
&\quad + \beta_{4}G_{it} + \beta_{5}TG_{it} + \beta_{6}F_{it} + \beta_{7}EU_{it} \\
&\quad + \beta_{8}EPC_{it} + \gamma_{t} + u_{i} + \epsilon_{it}
\end{split}
\end{equation}

\begin{table}[!htb]
\centering
\caption{Comparison of Random Effects Models}
\label{tab:ModelComparison}
\begin{tabular}{lcc}
\toprule
Criteria & Model 1 & Model 2 \\
\midrule
R-Squared & 0.98216 & 0.98305 \\
Adj. R-Squared & 0.98187 & 0.98189 \\
Total Sum of Squares & \(5.37 \times 10^{12}\) & \(5.36 \times 10^{12}\) \\
Residual Sum of Squares & \(9.59 \times 10^{10}\) & \(9.09 \times 10^{10}\) \\
AIC & 10972.74 & 10994.36 \\
BIC & 11010.67 & 11133.44 \\
Chisq & 27031.7 & 27087.8 \\
DF & 8 & 32 \\
p-value & \(< 2.22 \times 10^{-16}\) & \(< 2.22 \times 10^{-16}\) \\
\bottomrule
\end{tabular}
\end{table}

Although both models exhibit commendable data fit, the second model, with its marginally superior R-squared and lower residual sum of squares, suggests a nuanced understanding through the incorporation of a time trend. However, the increase in AIC and BIC values signals a caution against overcomplicating the model without clear justification. To address this, we have employed the Wald test to evaluate the following hypotheses:
\begin{itemize}
    \item Null Hypothesis (\(H_0\)): The coefficients of time-specific effects are jointly zero, indicating the extended model does not significantly outperform the baseline model.
    \item Alternative Hypothesis (\(H_1\)): At least one time-specific effect coefficient is nonzero, suggesting the second model provides a superior fit.
\end{itemize}

Given an F-test value of 1.0544 and a p-value of 0.394, we fail to reject $H_0$, indicating no substantial evidence that time-specific effects are significantly contributing to the model. This outcome favors the simpler random-effects model without the time trend for its parsimony and effective representation of the data. The results of this simpler random-effects model are presented in Table~\ref{tab:regression_results}.

\begin{table}[!htb]
\centering
\caption{Random Effects Model Results}
\label{tab:regression_results}
\begin{tabular}{lccccc}
\toprule
Term & Estimate & Std. Error & z-value & p-value & Significance \\
\midrule
Intercept & -62514.000 & 22289.000 & -2.8046 & 0.005037 & ** \\
RE & 462.66000 & 302.8000 & 1.5279 & 0.126534 & \\
Fa & 658.13000 & 458.42000 & 1.4356 & 0.151103 & \\
EPF & 247.48000 & 157.84000 & 1.5679 & 0.116907 & \\
G & 842.2000 & 305.46000 & 2.7571 & 0.005831 & ** \\
TG & 0.920190 & 0.006387 & 144.0800 & $<$ 2.2e-16 & *** \\
F & -0.092565 & 0.012098 & -7.6515 & 1.987e-14 & *** \\
EU & 1.807400 & 2.331900 & 0.7751 & 0.438292 & \\
EPC & -0.717160 & 0.677250 & -1.0589 & 0.289628 & \\
\bottomrule
\end{tabular}
\end{table}

Table~\ref{tab:regression_results} highlights key findings regarding the significance of various characteristics in the context of CO2 emission. Notably, GDP, total greenhouse gas emission, and forest area emerge as significant predictors, each demonstrating a substantial role at the 0.05 level, with p-values of 0.005, \(2.2 \times 10^{-16}\), and \(1.987 \times 10^{-14}\), respectively. Other variables, showing p-values ranging from 0.11 to 0.43, suggest their non-significant impact at the conventional 5\% significance level. This analysis prompts a focus on fixed effect models to examine entity-specific variations on CO2 emission, contrasting fixed characteristics against random, unobserved factors.

\subsubsection{Fixed Effects Models}

This subsection explores the application of two distinct fixed effects models to our dataset. The first model implements a linear regression framework augmented by entity-specific fixed effects. This adjustment allows for the accommodation of unobserved heterogeneity among entities by introducing unique intercepts for each entity. The second model employs a generalized linear model (GLM) framework, assuming a Gaussian distribution. This approach mirrors the structure of the first model but is designed to offer enhanced robustness in the presence of heteroscedasticity or non-normal error distributions. The equations governing these models are as follows:

\begin{equation}
\begin{split}
C_{oit} &= \alpha_{i} + \beta_{1}RE_{it} + \beta_{2}Fa_{it} + \beta_{3}EPF_{it} + \beta_{4}G_{it} \\
&\quad + \beta_{5}TG_{it} + \beta_{6}F_{it} + \beta_{7}EU_{it} + \beta_{8}EPC_{it} + \epsilon_{it}
\end{split}
\end{equation}

In the above equation, \(\alpha_i\) symbolizes the entity-specific fixed effect for entity \(i\), \(\beta\) values denote the coefficients of the respective predictors, and \(\epsilon_{it}\) is the error term. Models D and E, while sharing the same structural formula, diverge in their estimation techniques and foundational assumptions. Model D employs dummy variables for entity-specific fixed effects, offering precise capture of entity variations. Model E, within a GLM framework and assuming a Gaussian distribution, seeks robustness under specific conditions. The choice between fixed and random effects models is guided by a Hausman test, with hypotheses:

\begin{itemize}
    \item \(H_0\): The random effects estimator is consistent and efficient, implying no correlation between individual-specific effects and predictors.
    \item \(H_1\): The random effects estimator is inconsistent, favoring the fixed effects estimator due to correlations of individual-specific effects with predictors.
\end{itemize}

Rejecting \(H_0\) (with p-value $< 0.05$) favors fixed effects models for our analysis. Comparisons based on AIC and BIC indicate Model E as the superior fit, underscored by its lower BIC and log-likelihood, indicating a close alignment between model predictions and actual data.

\begin{table}[!htb]
\centering
\caption{Fixed Effects Model Results}
\label{tab:regression_results8}
\begin{tabular}{p{0.20\textwidth}ccccc}
\toprule
\textbf{Variable} & \textbf{Estimate} & \textbf{Standard Error} & \textbf{t-value} & \textbf{p-value} & \textbf{Significance} \\
\midrule
Renewable energy consumption & \(7.40 \times 10^{2}\) & \(2.90 \times 10^{2}\) & 2.5518 & 0.011 & * \\
Forest area (\% of land area) & \(-2.64 \times 10^{2}\) & \(1.50 \times 10^{3}\) & -0.1761 & 0.8603 & \\
Electricity production from oil, gas, and coal sources & \(3.73 \times 10^{2}\) & \(1.53 \times 10^{2}\) & 2.4337 & 0.015 & * \\
GDP per unit of energy use & \(8.74 \times 10^{2}\) & \(3.51 \times 10^{2}\) & 2.4928 & 0.013 & * \\
Total greenhouse gas emission & \(9.63 \times 10^{-1}\) & \(7.54 \times 10^{-3}\) & 127.77 & 0.0001 & *** \\
Forest area (sq. km) & \(-5.46 \times 10^{-1}\) & \(9.73 \times 10^{-2}\) & -5.616 & 0.0001 & *** \\
Energy use & \(-1.95 \times 10^{0}\) & \(2.32 \times 10^{0}\) & -0.843 & 0.3994 & \\
Electric power consumption & \(3.60 \times 10^{-2}\) & \(6.56 \times 10^{-1}\) & 0.055 & 0.9563 & \\
\bottomrule
\end{tabular}
\end{table}

The examination of variable significance as shown in Table~\ref{tab:regression_results8}, with a 0.05 level set for determining impactful variables, reveals Renewable Energy Consumption, Electricity Production from oil, gas, and coal sources, and GDP per Unit of Energy Use as significantly positively impacting the dependent variable, marked by p-values of 0.01103, 0.01531, and 0.01301, respectively. Total Greenhouse Gas emission emerge as the most influential variable, exhibiting a strong positive correlation with a p-value of less than 0.0001. Conversely, Forest Area (sq. km) significantly negatively affects the dependent variable, also with a p-value of less than 0.0001. Variables such as Forest Area (\% of land area), Energy Use, and Electric Power Consumption did not achieve statistical significance at the 0.05 level, as indicated by their higher p-values. Given the high correlation coefficient (-0.83 as depicted in Figure~\ref{fig:myimage5}) between Renewable energy consumption and Electricity production from oil, gas, and coal sources, and considering the lower p-value associated with Renewable energy consumption,we have only incorporated the former variable in subsequent analyses. This decision is aimed at refining our examination of predictors while proactively addressing concerns regarding multicollinearity.
This comprehensive approach to model selection and analysis underscores the importance of considering both model fit and the statistical significance of individual variables, guiding us toward a robust understanding of the determinants of CO2 emission. 

\subsection{Phase II: Prediction and Cluster Analysis of CO2 emission}

This phase analyzes and predicts CO2 emission across nations by utilizing the features identified in Phase I. This section focuses on predicting emission over a three-year period. Following these predictions, countries are clustered by their CO2 emission trends, and the clusters have been then analyzed to gain a deeper understanding of different emission patterns using the most significant indicators identified in Phase I.

\subsubsection{Prediction of CO2 emission}

\begin{figure}[!htb]
\centering
\includegraphics[width=0.95\linewidth]{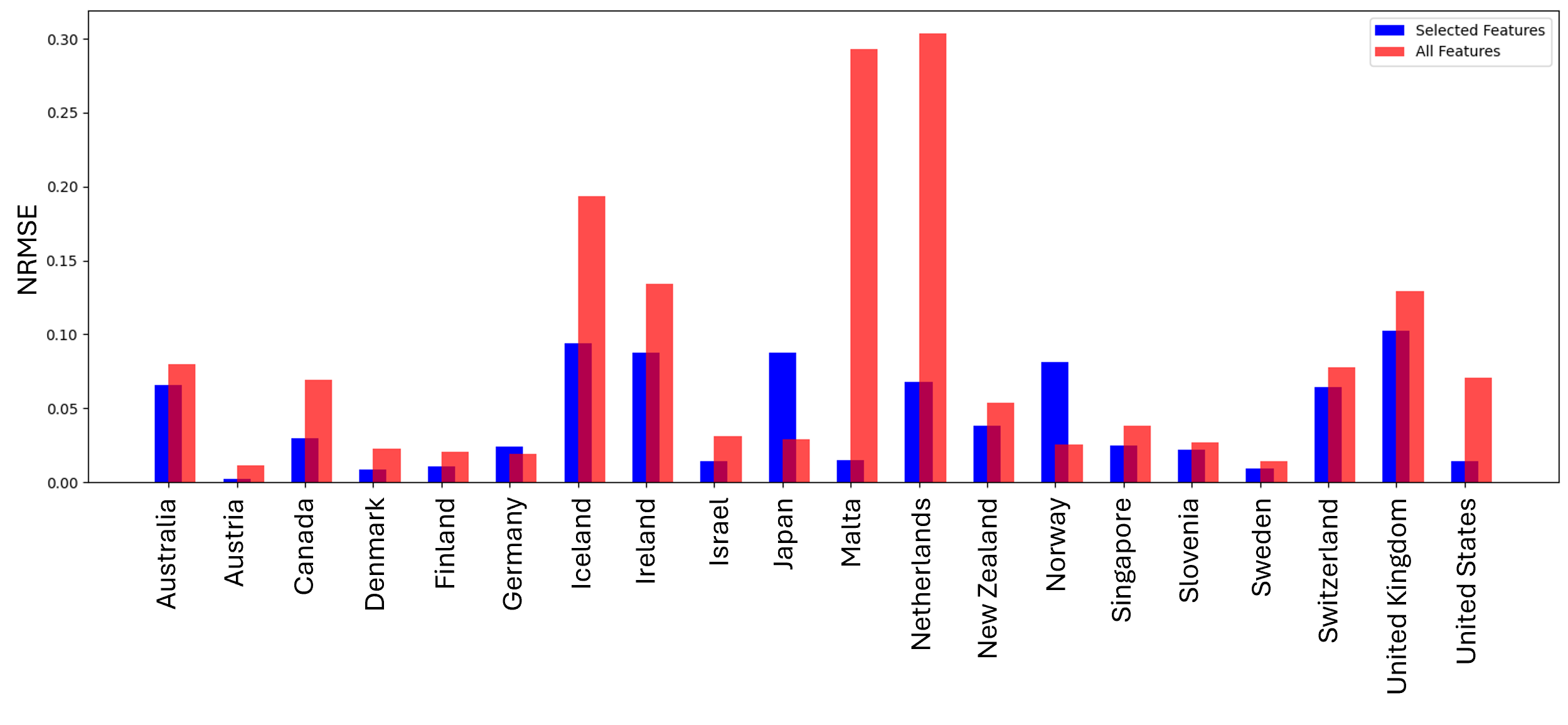}
\caption{NRMSE comparison across countries using all features versus selected features from Phase I}
\label{fig:myimage10}
\end{figure}

Based upon the key characteristics identified in Phase I, such as renewable energy consumption, GDP per unit of energy use, greenhouse gas emission, and forest area, CO2 emission are predicted from 2012 to 2014. The prediction of CO2 emission is performed using the SARIMAX model where prediction results from two different scenarios are presented. In the first scenario, we predict CO2 emission over a three-year period using all the 8 features of CO2 indicators. In the second scenario, the prediction with SARIMAX over the same three-year period is done using the 4 selected features that we have obtained from Phase I implementing the best model (Model E). Table 9 highlights the RMSE and MAE of the SARIMAX model for both these scenarios, whereas Figure~\ref{fig:myimage10} illustrates the NRMSE across different countries, comparing the outcomes using all the available features with the features selected from Phase I.

\begin{table}[!htb]
\centering
\caption{RMSE and MAE of SARIMAX model predictions using all features versus selected features from Phase I}
\label{tab:feature_comparison}
\begin{tabular}{lcccc}
\toprule
Country & \multicolumn{2}{c}{All Features} & \multicolumn{2}{c}{Selected Features} \\
\cmidrule(lr){2-3} \cmidrule(lr){4-5}
 & RMSE & MAE & RMSE & MAE \\
\midrule
Switzerland & 435.46 & 433.32 & 298.27 & 277.25 \\
Norway & 321.19 & 318.40 & 1036.16 & 903.42 \\
Iceland & 95.76 & 95.59 & 46.47 & 37.64 \\
Australia & 24051.84 & 23682.12 & 8693.57 & 8692.87 \\
Denmark & 699.83 & 541.63 & 250.14 & 242.44 \\
Sweden & 328.78 & 267.76 & 189.28 & 183.01 \\
Ireland & 1641.91 & 1444.42 & 1428.17 & 1401.92 \\
Germany & 4612.22 & 4328.02 & 5268.25 & 4508.37 \\
Netherlands & 6716.53 & 6069.93 & 1501.10 & 1382.38 \\
Finland & 398.62 & 374.73 & 199.88 & 190.58 \\
Singapore & 648.48 & 586.43 & 389.72 & 334.27 \\
New Zealand & 793.51 & 676.55 & 456.41 & 435.40 \\
Canada & 10962.22 & 10893.59 & 4654.82 & 3003.13 \\
United Kingdom & 16159.76 & 13459.35 & 12826.57 & 9421.95 \\
Japan & 9982.56 & 7788.30 & 12081.59 & 11135.11 \\
Israel & 1284.12 & 1122.49 & 513.74 & 424.41 \\
United States & 68695.06 & 61815.72 & 13433.16 & 12225.59 \\
Malta & 206.78 & 157.41 & 10.28 & 7.77 \\
Slovenia & 170.08 & 144.07 & 101.13 & 90.83 \\
Austria & 209.25 & 197.29 & 41.04 & 36.72 \\
\bottomrule
\end{tabular}
\end{table}

The results presented in Table \ref{tab:feature_comparison} indicate that the SARIMAX model benefits significantly from utilizing selected features from Phase I in predicting CO2 emission. Switzerland, in particular, experienced a 31.5\% decrease in RMSE and a 36.0\% decrease in MAE, demonstrating the effectiveness of the feature selection. In case of Australia, RMSE and MAE decreased to 8693.57 and 8692.87, respectively, from 24051.84 and 23682.12. These improvements suggest that better emission estimates can be achieved by using a more focused set of features. As shown in Figure~\ref{fig:myimage10}, the NRMSE across various countries exhibits significant reductions when the selected features are applied. In Norway, the NRMSE decreased by approximately 50\%, which indicates a significant improvement in the accuracy of the model. Although the selected features did not yield uniformly lower NRMSE values for all countries, as was observed with Germany and Japan, the overall trend indicates that the strategic selection of features can contribute to significant improvements in emission prediction, thus supporting the research hypothesis.

\begin{figure}[!htb]
\centering
\includegraphics[width=0.95\linewidth]{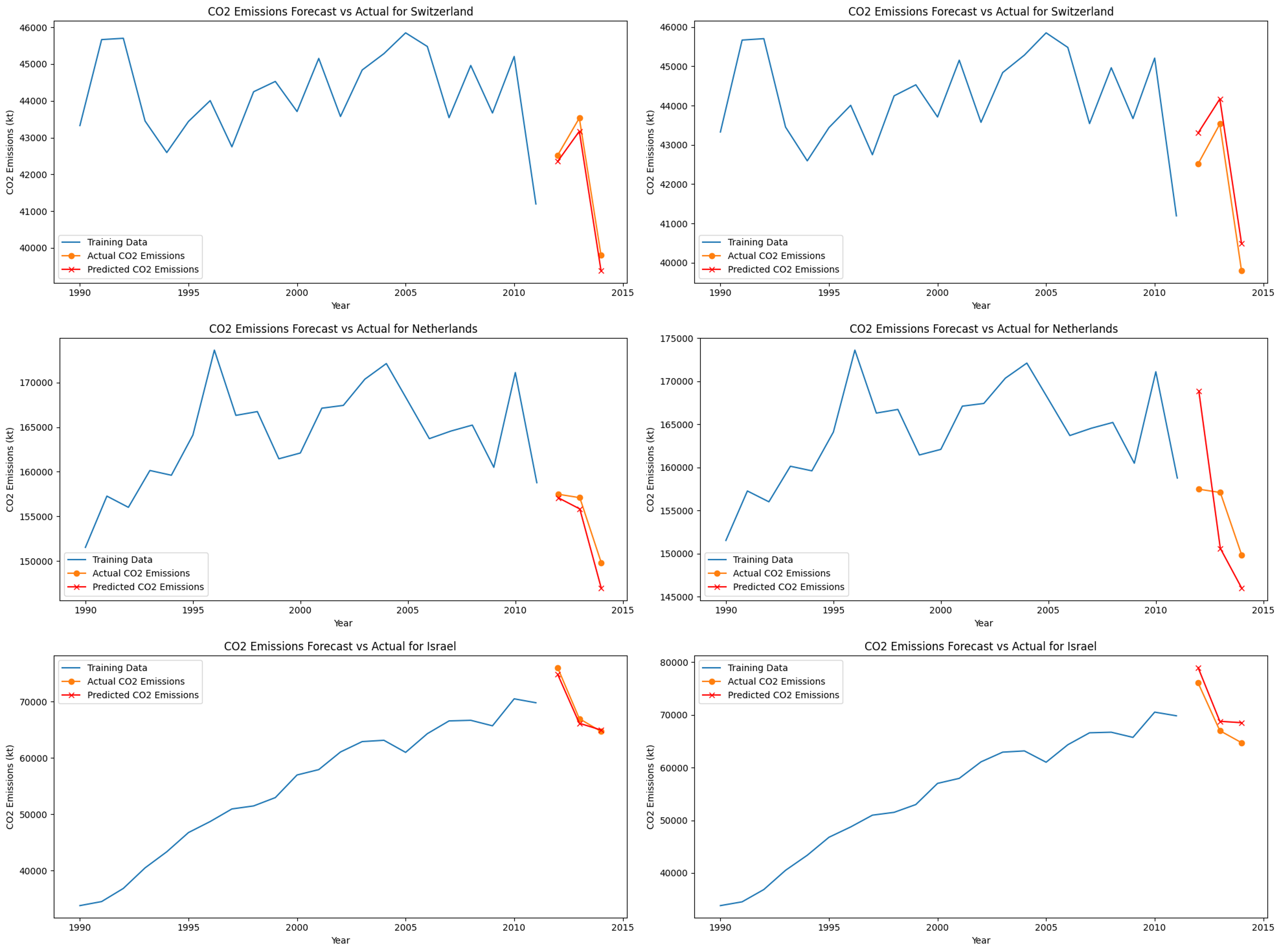}
\caption{SARIMAX model predictions for three representative countries using selected features from Phase I (left) and all features (right)}
\label{fig:myimage11}
\end{figure}

By examining three representative countries categorized according to their relative HDI positions, we can gain a deeper understanding of the model's predictive performance. In Figure~\ref{fig:myimage11}, we compare the model's performance using the entire set of features against the predictions made with the selected features from Phase I. Based on this comparison, it appears that using selected features generally enhances the model's prediction performance, as demonstrated by its closer approximation to actual CO2 emission for these countries. An in-depth examination of the Netherlands' predictive results provides valuable insight into the approach's effectiveness. SARIMAX, using its entire feature set, incorrectly projects an increase in CO2 emission from 2011 to 2012. However, this is in contradiction to the actual emission trajectory, which indicates a decline. In contrast, the model's prediction is more closely aligned with the actual emission data when the selected features from Phase I are applied, which accurately reflects the downward trend in emission. This finding emphasizes the effectiveness of careful feature selection in improving emission prediction accuracy.

\subsubsection{Cluster Analysis}

\begin{figure}[!htb]
\centering
\includegraphics[width=0.95\linewidth]{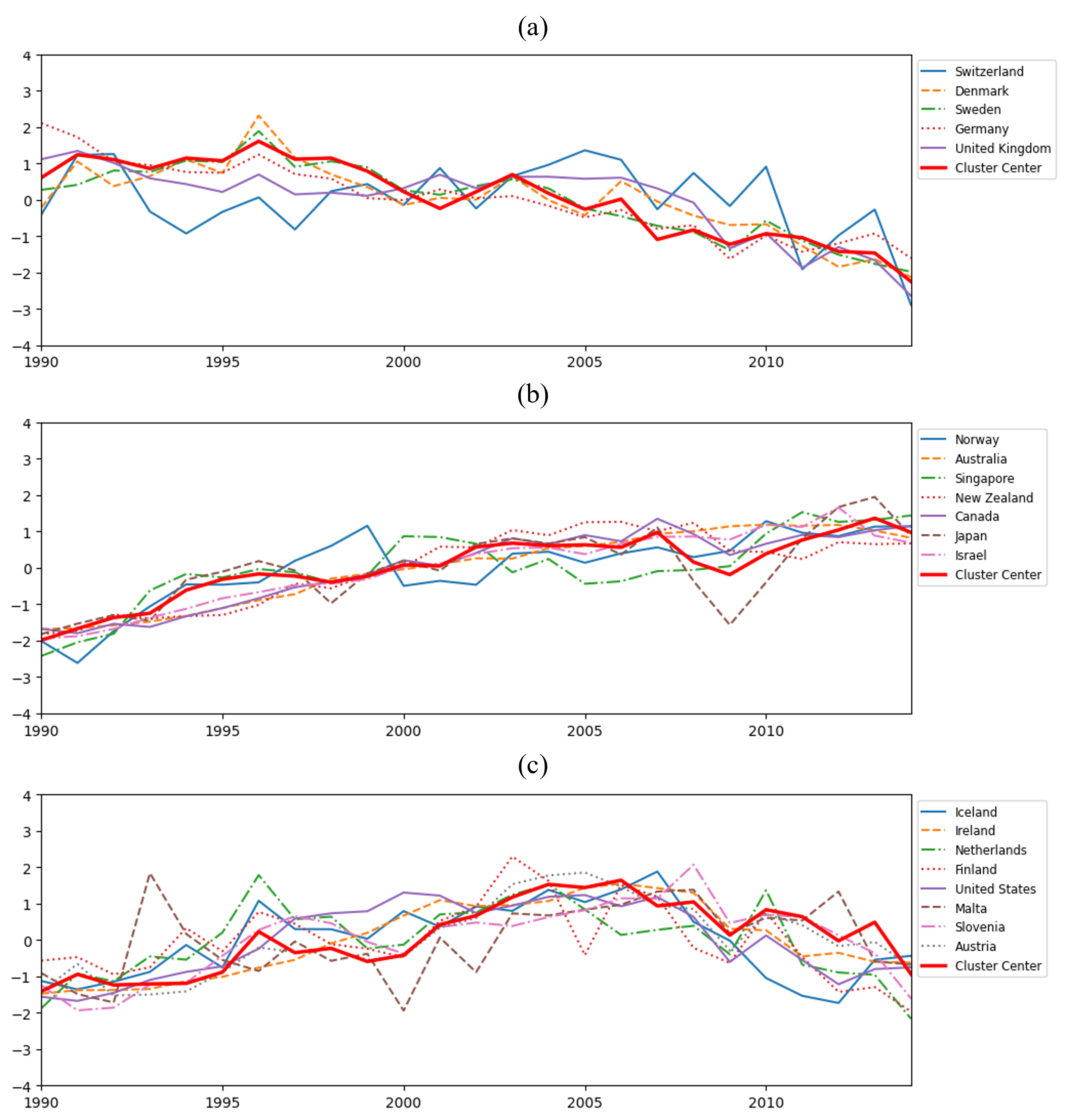}
\caption{Cluster analysis of the countries based on carbon emission patterns}
\label{fig:myimage12}
\end{figure}

After predicting CO2 emission, in this subsection, we perform a cluster analysis by applying Dynamic Time Warping (DTW) algorithm which categorizes the countries based on their CO2 emission trends. This methodology, which builds on insights obtained in the data understanding phase, effectively groups countries with similar emission patterns into three clusters as shown in Figure~\ref{fig:myimage12}.

\begin{figure}[!htb]
\centering
\includegraphics[width=0.95\linewidth]{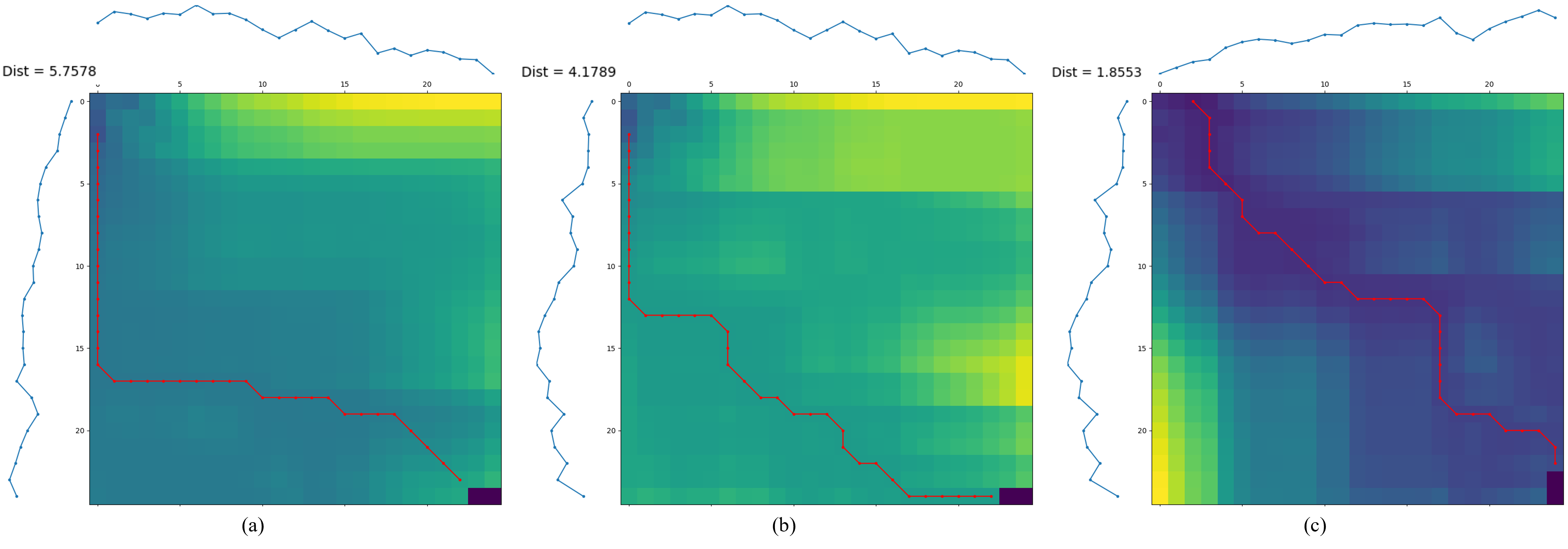}
\caption{Dynamic Time Warping distances between cluster centers for CO2 emission}
\label{fig:myimage13}
\end{figure}

\begin{figure}[!htb]
\centering
\includegraphics[width=0.95\linewidth]{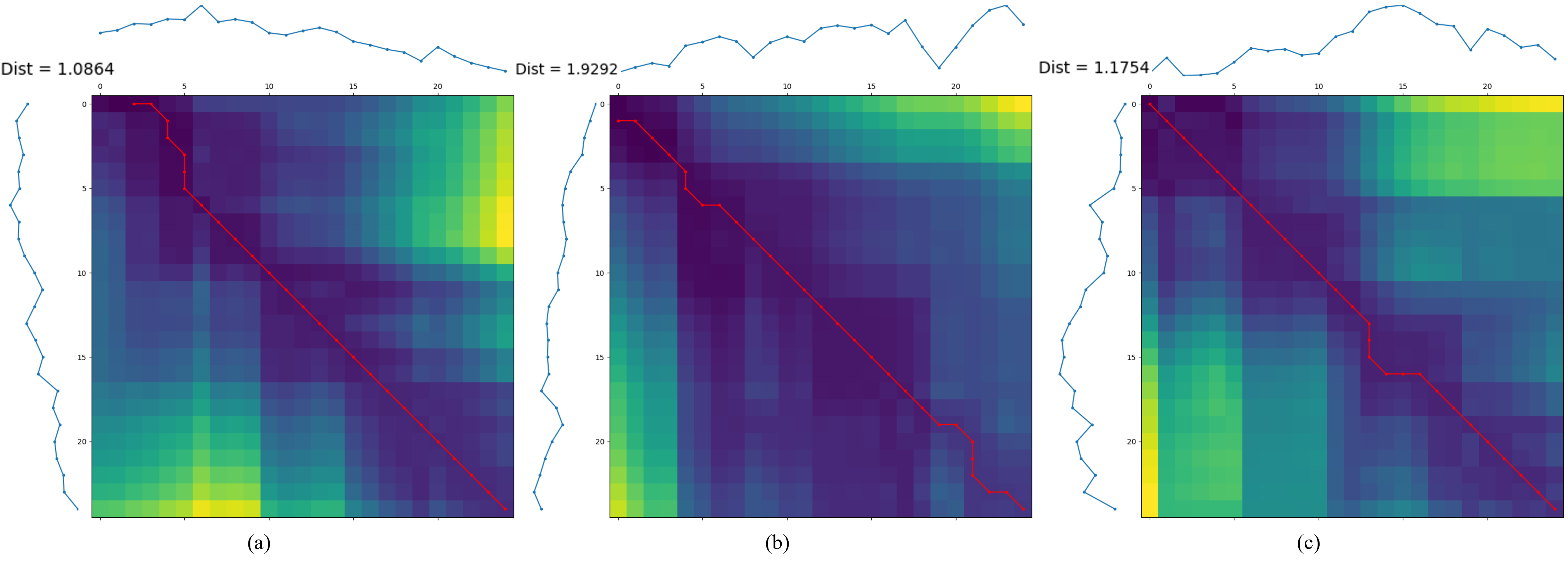}
\caption{Alignment of individual country CO2 emission trends with cluster central trends}
\label{fig:myimage14}
\end{figure}

In Figure~\ref{fig:myimage12}, along x-axis we have time spanning from the year 1990 to 2014, and the y-axis indicates the standardized CO2 emission levels for each country. By standardizing the emission trends, it is possible to observe patterns and dynamics of CO2 emission for each country more easily. The first cluster, as shown in Figure~\ref{fig:myimage12}(a) includes countries such as Switzerland and the United Kingdom, whose emission patterns exhibit notable fluctuations but tend to converge in later years towards the cluster center. In addition to having a variety of industrial activities, this group may also reflect economies that have progressively adopted emission reduction strategies. As illustrated in the Figure~\ref{fig:myimage12}(b), Norway, Australia, and Canada are included in the second cluster. These nations share a consistent upward trend in CO2 emission before stabilizing or slightly decreasing as they approach later times. This pattern may indicate sustained growth in emission caused by economic expansion, followed by a shift toward stabilization, possibly influenced by environmental policies. A third and final cluster can be seen in Figure~\ref{fig:myimage12}(c), which includes the United States and the Netherlands, among other countries. A trend of increasing emission is observed in this diverse group until 2004, following which a gradual decline is observed. These trends may be associated with a peak in industrial emission followed by concerted efforts to reduce emission and adopt renewable energy sources. During the period examined, there were distinct trajectories in the emission of CO2 in these countries, which may reflect the different socioeconomic and environmental policies of each country. A comprehensive evaluation of the robustness of our clustering approach is provided by Figures 12 and 13 which provide insight into how CO2 emission trends are aligned. In Figure~\ref{fig:myimage13}, the distances between cluster centers are presented, revealing inter-cluster variability, whereas in Figure~\ref{fig:myimage14}, an individual country trend is compared to the central trend of the cluster, revealing intra-cluster consistency.

\begin{table}[!htb]
\centering
\caption{Analysis of the clusters based on the selected features}
\label{tab:features_summary}
\begin{tabular}{p{3cm}ccccp{1.5cm}}
\toprule
\textbf{Feature} & \textbf{Cluster} & \textbf{Mean} & \textbf{Median} & \textbf{Standard Deviation} & \textbf{Annual Growth Rate (\%)} \\
\midrule
\multirow{3}{=}{Renewable energy consumption} & Cluster 1 & 15.95 & 13.64 & 13.57 & -3.86 \\
 & Cluster 2 & 18.1 & 8.42 & 18.91 & -10.91 \\
 & Cluster 3 & 18.78 & 9.73 & 20.52 & -1.76 \\
\midrule
\multirow{3}{=}{Electricity production from oil, gas and coal sources} & Cluster 1 & 43.37 & 61.03 & 34.41 & 17.12 \\
 & Cluster 2 & 57.33 & 61.18 & 37.66 & 35.07 \\
 & Cluster 3 & 54.67 & 53.39 & 33.2 & 22.61 \\
\midrule
\multirow{3}{=}{GDP per unit of energy use} & Cluster 1 & 8.65 & 8.12 & 3.6 & 2.7 \\
 & Cluster 2 & 6.95 & 6.34 & 2.97 & 5.21 \\
 & Cluster 3 & 6.88 & 6.23 & 3.53 & 6.93 \\
\midrule
\multirow{3}{=}{Total greenhouse gas emission} & Cluster 1 & 364635.94 & 73846.74 & 385349.73 & 9.94 \\
 & Cluster 2 & 389649.8 & 83368.13 & 432138.17 & 2.59 \\
 & Cluster 3 & 855195.2 & 69949.81 & 2102668.71 & 13.84 \\
\midrule
\multirow{3}{=}{Forest area} & Cluster 1 & 88400.66 & 29750.0 & 104200.41 & 4.26 \\
 & Cluster 2 & 751927.56 & 121149.0 & 1195835.76 & -16.44 \\
 & Cluster 3 & 417152.51 & 9722.06 & 1001584.13 & 25.37 \\
\bottomrule
\end{tabular}
\end{table}

Dynamic Time Warping (DTW) visualizations provide a robust analytical tool for validating CO2 emission trends clustering across various countries. Figures~\ref{fig:myimage13}(a), ~\ref{fig:myimage13}(b), and~\ref{fig:myimage13}(c) demonstrate the temporal alignment between cluster centers, illustrating disparities between clusters. In each heatmap, the x-axis shows the sequential time step corresponding to the standardized emission values for one cluster, while the y-axis relates to the time step corresponding to another cluster. In these figures, a gradient of colors indicates local point-to-point distances, with warmer colors indicating higher distance and cooler colors showing closer proximity, which illustrates the degree of similarity between the two time series. There is notable variation in the optimal warping paths represented by the red lines, with deviations from the diagonal indicating periods of divergence in CO2 emission trends between the clusters.

Additionally, Figures~\ref{fig:myimage14}(a),~\ref{fig:myimage14}(b), and~\ref{fig:myimage14}(c) examine the alignment between emission trends within individual countries and cluster centers, thus examining the coherence of countries within their clusters. Each heatmap's x-axis represents the central trend of the cluster, while its y-axis corresponds to the central trend of a particular country within that cluster. A strong alignment with the cluster's central trend is indicated by the proximity of the red lines to the diagonal, which indicates that the optimal warping paths are aligned. Overall DTW distances at the top of each heatmap provide a quantitative measure of the alignment, and the small distances indicate that the clustering is reliable. In the Table \ref{tab:features_summary}, different clusters are analyzed based on the selected features from Phase I.

As indicated by the moderate and declining growth rate of renewable energy consumption in Cluster 1, which can be referred to as the ``Reducing emission Group'', CO2 emission were decreasing. Despite the significant increase in the annual growth rate of electricity produced by oil, gas, and coal, the mean electricity production from these sources is 44.37 percent, suggesting a transition towards cleaner energy sources has occurred. A high GDP per unit of energy use, as well as a growth in forest area, support the hypothesis that the countries belonging to this cluster have moved towards sustainable practices. As a result of an increase in CO2 emission, Cluster 2 can be identified as a group with ``Growing emission". This is consistent with the high mean of 57.33\% in electricity production from oil, gas, and coal sources and the highest growth rate of 35.07\%, which indicates a heavy reliance on non-renewable energy sources. Although the cluster is characterized by the largest forest area, its negative annual growth rate is concerning, potentially exacerbating its overall environmental impact. In Cluster 3, the ``Peak and Decline emission Group", CO2 emission initially increased, followed by a decline, resulting in the largest mean renewable energy consumption at 18.78\% and a less pronounced decrease in growth rate. As indicated by the highest annual growth rate in forest area, which is 25.37 percent, significant total greenhouse gas emission are aligned with this pattern but are attempting to reverse the trend. A summary of the results over the period from 1990 to 2014 suggested that Cluster 1 had made significant progress toward reducing emission. Cluster 2 experienced an upward trend in emission, indicating a growing environmental challenge during this period. Meanwhile, Cluster 3 appeared to correct its initially high emission, demonstrating a shift towards more sustainable practices in the latter part of the timeframe. These historical trends, analyzed in conjunction with the detailed characteristics identified in Phase I, provided a comprehensive perspective on the environmental trajectories of the clusters and the effectiveness of their energy and environmental policies.

\section{Conclusion}

In the context of addressing global climate change, this research marks a significant step forward by pinpointing the factors that drive CO2 emission. By combining a broad range of variables with diverse statistical models, the study provides an in-depth and comprehensive analysis. Its standout feature is a two-phase methodology that focuses on analyzing key features and enhancing prediction precision, leading to notably improved forecasting accuracy. This approach ensures predictions mirror actual trends in CO2 emission, whether they're rising or falling, thus making future projections more dependable. A key methodological breakthrough of this research is the application of the Dynamic Time Warping (DTW) algorithm during its second phase, enabling the categorization of countries into clusters based on their emission trajectories. This provides deep insights into the global landscape of CO2 emission, highlighting both the commonalities and disparities across nations.

This study has substantial practical implications, especially pertaining to policy development. By providing precise forecasts of future emission trends and a detailed understanding of emission dynamics on a country-by-country basis, it empowers policymakers and environmental strategists to devise climate policies that are both impactful and finely tuned to specific needs. Thus, the study's contributions extend far beyond academic circles, offering practical tools for combatting climate change. It aids in informed policymaking and strategic environmental planning, significantly aiding the global initiative to curb CO2 emission and counteract climate change's negative impacts.

\backmatter



\vspace{10 pt}
\noindent\textbf{Funding} This research has been generously supported by the United States Environmental Protection Agency (EPA) under the Pollution Prevention (P2) practices grant.

\noindent\textbf{Data availability} Data will be available upon reasonable request to authors.

\noindent\textbf{Author Contributions} Conceptualization: Imtiaz Ahmed, Hamed Khosravi, Farzana Islam; Methodology: Hamed Khosravi, Farzana Islam,
 Imtiaz Ahmed; Formal analysis and investigation: Hamed Khosravi, Farzana Islam, Ahmed Shoyeb Raihan; Writing—original draft preparation: Hamed Khosravi, Farzana Islam, Ahmed Shoyeb Raihan; Writing—review and editing:
 Ashish Nimbarte, Imtiaz Ahmed; Supervision: Ashish Nimbarte, Imtiaz Ahmed.

\section*{Declarations}
\begin{itemize}
\item \textbf{Competing interests} The authors declare no competing interests.
\item \textbf{Consent to participate} All authors have consent to participate.
\item \textbf{Consent for publication} All authors have consent for publication.
\item \textbf{Ethical Approval} Not Applicable.
\end{itemize}

\bibliography{sn-bibliography}

\end{document}